\PassOptionsToPackage{dvipsnames}{xcolor}
\documentclass[aps,prx,reprint,groupedaddress]{revtex4-2}

\usepackage{mathtools, nccmath}
\usepackage{amssymb}
\usepackage{dsfont}
\usepackage{afterpage}
\usepackage{float}
\usepackage{pgfplots}
\usepackage{tikz}
\usetikzlibrary{decorations.pathmorphing}
\usetikzlibrary{arrows, backgrounds}
\usepackage{ifthen}
\usepackage{placeins}
\usepackage{graphicx}
\graphicspath{ {paper_fig/} }
\definecolor{almond}{rgb}{0.94, 0.87, 0.8}

\newcommand{\be}{\begin{equation}}
\newcommand{\ee}{\end{equation}}
\newcommand{\bea}{\begin{eqnarray}}
\newcommand{\eea}{\end{eqnarray}}
\newcommand{\Eq}[1]{Eq.\,(\ref{#1})}
\newcommand{\Eqs}[2]{Eqs.\,(\ref{#1}) and (\ref{#2})}
\newcommand{\Eqsss}[3]{Eqs.\,(\ref{#1}), (\ref{#2}), and (\ref{#3})}

\newcommand{\Fig}[1]{Fig.\,\ref{#1}}
\newcommand{\Figs}[2]{Figs.\,\ref{#1} and \ref{#2}}

\newcommand{\Figsss}[3]{Figs.\,\ref{#1}, \ref{#2}, and \ref{#3}}
\newcommand{\Sec}[1]{Sec.\,\ref{#1}}
\newcommand{\App}[1]{Appendix\,\ref{#1}}
\newcommand{\Tab}[1]{Tab.\,\ref{#1}}

\newcommand{\rmand}{\quad{\rm and}\quad}

\begin{document}

\title{F\"orster transfer between quantum dots in a shared phonon environment: An exact approach, revealing the role of pure dephasing}

\author{L. S. Sirkina}
\email[]{sirkinaluba@gmail.com}
\affiliation{School of Physics and Astronomy, Cardiff University, Cardiff CF24 3AA, United Kingdom}

\author{L.M.J. Hall}
\affiliation{School of Physics and Astronomy, Cardiff University, Cardiff CF24 3AA, United Kingdom}

\author{A. Morreau}
\affiliation{School of Physics and Astronomy, Cardiff University, Cardiff CF24 3AA, United Kingdom}

\author{W. Langbein}
\affiliation{School of Physics and Astronomy, Cardiff University, Cardiff CF24 3AA, United Kingdom}

\author{E. A. Muljarov}
\affiliation{School of Physics and Astronomy, Cardiff University, Cardiff CF24 3AA, United Kingdom}

\date{\today}

\begin{abstract}

F\"orster resonance energy transfer has an important role in nature and technology, rendering its exact theoretical understanding significant. To this end, a system of two electronically decoupled quantum dots (QDs) is considered, interacting via dipole-dipole interaction and a common phonon bath. While the former leads to an oscillatory excitation transfer between the dots, the latter provide the dissipation resulting in directional F\"orster transfer. We present an exact microscopic treatment of the phonon-assisted transitions between hybridized exciton levels of the coupled QD system, going beyond the simple perturbative approaches commonly used in the literature. From our asymptotically exact results we extract population decay times $T_1$, dephasing times $T_2$, and resulting pure dephasing times $T_2^*$ of the states. We compare this treatment with an analytical model based on Fermi's golden rule, combining the most accurate elements of existing analytical treatments. The exact results show a significant deviation from this model in some parameter regimes, mainly due to the role of multi-phonon processes, which become important for comparable electron-phonon and dipolar coupling, realised at short distances between the QDs and at elevated temperatures.
\end{abstract}

\maketitle

\section{Introduction}
F\"orster resonance energy transfer, originating from the exchange Coulomb interaction, is a purely quantum effect having both fundamental and practical importance. The indirect mechanism of the F\"orster transfer has a number of promising applications, including light harvesting~\cite{herek2002quantum,ishizaki2012quantum}, photonic logic gates~\cite{Claussen2014complex}, and sensing~\cite{Rajput2022fret}.
An improved understanding of the physics of the F\"orster transfer is also key for further optimisation of light harvesting technology. F\"orster transfer furthermore has various applications in life sciences~\cite{dos2020quantum}, such as biosensing and measuring molecular distances. From the fundamental point of view, the study of the phonon-assisted F\"orster mechanism can aid our understanding of photosynthesis on a quantum-mechanical level, and may spark further progress in biophysical applications. This will also help to understand other forms of resonant energy transfer that have recently been proposed, such as polariton-assisted~\cite{du2018theory,Saez2018organic} and plasmon-induced~\cite{wang2016molecular} mechanisms.

The first experimental observations of F\"orster transfer phenomenon were made in excited atoms in a mixture of mercury and hydrogen vapors~\cite{cario1922zerlegung}.
The theory of this mechanism was then gradually developed but it was first properly quantified by F\"orster in terms of the overlap integral between donor emission and acceptor absorption spectra~\cite{forster1948zwischenmolekulare}, providing a connection between theory and experiment. This theory was extended by Dexter~\cite{dexter1953theory} to include forbidden quadrupole transitions, which become relevant over smaller transfer distances. The resulting theory was very influential and remains popular to this day.

In semiconductor quantum dots (QDs), which are inevitably coupled to a phonon environment, the excitation transfer can be based on electronic tunnel-coupling or Coulomb interaction. If the QDs are electronically coupled, as in a QD molecule, the QD exciton wave functions overlap by tunneling, and phonon-assisted transitions between the states provide dissipative transfer. This is called phonon-assisted tunneling in the literature. The effects of diagonal and non-diagonal coupling to phonons have been rigorously studied in such systems~\cite{Muljarov2005phonon}, focusing on the role of the Coulomb interaction, tunneling, and structural asymmetry. QD molecules were also addressed with a perturbative treatment~\cite{Wu2005charge}, which introduces a F\"orster-like interaction, called tunnel coupling in that work.
Experimental works also report on the transfer of excitation due to phonon-assisted tunneling~\cite{Heitz1998excitation,Tackeuchi2000dynamics,Rodt2003lateral,Ortner2005energy,Nakaoka2006direct,Chang2008nonresonant},
or due to F\"orster coupling~\cite{Unold2005optical,Gerardot2005photon,Kim2008experimental}.
Some theoretical studies accounted for both mechanisms in the context of optically driven QD molecules, while also including biexcitonic effects~\cite{rolon2008forster, rolon2009forster}, but these were not considering phonons. Several other works also incorporate biexcitonic effects, while omitting the phonon contribution~\cite{Specht2015twodimensional,Lovett2003optical,Danckwerts2006theory}.

The treatment of F\"orster transfer between QDs has two limiting regimes which allow simplifications. One is a dipolar coupling between the excitons much weaker than the exciton-phonon interaction, so that the transfer process can be seen as incoherent jumps within the F\"orster phenomenology~\cite{forster1948zwischenmolekulare} and Fermi's golden rule (FGR)~\cite{dirac1927quantum}, widely used in the literature for calculation of transition rates in the F\"orster transfer~\cite{govorov2005spin,Thilagam2008decoherence}. Clearly, this approach neglects any effects of coherence or pure dephasing.
The other, opposite regime is the a coupling to the phonon bath much weaker than the dipolar coupling, which can be addressed by standard approaches using Lindblad~\cite{lindblad1976generators} or Redfield~\cite{redfield1965theory} theories. In this regime, Lindblad~\cite{rolon2008forster, rolon2009forster}, Redfield~\cite{Novoderezhkin2003Exciton,rouse2023light}
and other approaches~\cite{Pachon2011Physical} have been applied to address energy transfer, but this picture does not provide insight into the role of the environment in the energy transfer process.
In the intermediate regime when the two couplings are comparable, which is of particular relevance in e.g. photosynthesis~\cite{Ishizaki2009adequacy}, both approximations fail. 
There have been works which modify the above approaches in order to address the intermediate regime, e.g. by combining Lindblad master equations with quantum walks~\cite{Mohseni2008Environment} or with FGR~\cite{leon2014importance}, or a pure dephasing rate is added phenomenologically as a tunable parameter, the latter receiving attention recently~\cite{Cortes2022fundamental,diaz2023composed}.

To properly address the intermediate regime, bridging the two regimes of weak dipolar coupling and weak coupling to the bath, other theoretical tools must be employed. Various methods have been suggested to address quantum coherence in the F\"orster transfer. These include
Markovian Bloch–Redfield theory~\cite{Thilagam2008decoherence,wu2010efficient,moix2011efficient},
hierarchical equations of motion~\cite{novoderezhkin2023excitation,wang2018efficient,
kramer2018efficient},
second-order correlation expansions~\cite{Machnikowski2008quantum},
second-order cumulant expansions~\cite{richter2006theory,
ma2015forster1,ma2015forster2,nalbach2011iterative},
polaron master equations~\cite{Nazir2009correlation,McCutcheon2011coherent,
jang2008theory,jang2011theory}
and other similar approaches~\cite{hwang2015coherent,Huo2011Communication,
Rivera2013Influence}. Some of these approaches were compared in~\cite{tao2020coherent,Pachon2012Computational} and reviewed in~\cite{Chin2012Coherence,chenu2015coherence}. The F\"orster transfer process so far has been studied more extensively for chromophores in proteins. Among the earlier mentioned works,  F\"orster transfer involving a pair of QDs in a shared phonon bath was considered in \cite{govorov2005spin,Thilagam2008decoherence,diaz2023composed,
Nazir2009correlation,McCutcheon2011coherent, Machnikowski2008quantum, richter2006theory}.

A more realistic treatment of F\"orster transfer in the intermediate regime requires a rigorous treatment of the bath which facilitates the transfer.
The properties of a shared bath have been linked~\cite{olaya2011energy} to long-lived superpositions of exciton states reported in two-dimensional photon echo spectroscopy experiments with proteins~\cite{engel2007evidence,lee2007coherence}, highlighting the inadequacy of Lindblad and Redfield treatments~\cite{Ishizaki2009adequacy}.
It has been shown that a shared bath can result in non-classical system-bath correlations~\cite{chen2019quantifying,chen2018simulating}.
However, a rigorous treatment of the bath can only be achieved by
path-integral based approaches~\cite{moix2015forster3,nalbach2011iterative,
kundu2022intramolecular,gribben2020exact}. Only recently such path-integral based approaches became available~\cite{morreau2019phonon, sirkina2023numeric, strathearn2018efficient}, opening up such an opportunity. However, these exact methods have not been applied to F\"orster transfer so far. Recently, a system of dipolar coupled QDs in a shared bath has been treated by a path-integral based approach~\cite{gribben2020exact}, but the F\"orster transfer itself was not reported, and phonons were limited to one dimension. While a realistic representation of exciton-phonon interaction in semiconductor QDs requires a three-dimensional treatment of phonons, thin disc-shaped QDs (or nanoplatelets~\cite{Naeem2015giant}) can approach the one-dimensional case.

In this paper, we present an exact microscopic calculation of the F\"orster transfer in QDs, which has not been available until now, and compare it with existing theories, demonstrating regimes of significant deviations. We consider the F\"orster transfer in a pair of spatially separated but coupled semiconductor QDs interacting with a continuum of tree-dimensional phonon modes, which constitute their common environment. The exchange of information and quantum excitation between the dots occurs both directly via a dipolar coupling and indirectly via the phonon environment. The latter is a source of optical decoherence in the QD pair and also facilitates the F\"orster transfer.

We implement an exact path integral-based approach, recently developed in Ref.~\cite{morreau2019phonon}. Using two key ingredients, Trotter's decomposition and linked cluster expansion, this approach provides asymptotically exact results even in regimes with comparable timescales of exciton-phonon and dipolar-coupling dynamics.  We take into account all higher-order multi-phonon processes and realistic three-dimensional phonons.

From these results, we derive an analytical approach, which combines the long-time behavior of the exact approach with hybridization of QD states and phonon-assisted transitions between them incorporated via FGR. This allows us to compare the exact solution with existing approaches, and analyze the population dynamics and characteristic decoherence times. This analysis reveals a crucial role of pure dephasing in the F\"orster transfer mechanism in the regime of comparable exciton-phonon and F\"orster couplings.

The paper is structured as follows. The treated system is introduced in \Sec{sec:system}, an exact path integral-based approach is described in \Sec{sec:theory}, the analytical approximation is provided in \Sec{sec:analytics}, and results are given in \Sec{sec:results}, followed by the conclusions in \Sec{sec:conc}. A set of appendices provide supporting details.

\section{Physical system}
\label{sec:system}

We consider a pair of QDs, in which the exchange of an optical excitation is driven by a dipolar coupling. In addition to this coupling, both dots interact with a common bath -- a continuum of three-dimensional phonon modes.
We assume that the distance between the dots is large enough to neglect electronic coupling, but  small compared to the phonon coherence length. The Hamiltonian of this system consists of two exactly solvable parts,
\begin{align}
    H=H_{\rm F}+H_{\rm IB}\,.
    \label{hamilt}
\end{align}
The first part,
\begin{align}
H_{\rm F}&=\sum_{j=1,2} \Omega_j d^\dagger_j d_j+g_{\rm F}\left( d^{\dagger}_1 d_2 + d^{\dagger}_2 d_1\right)\,,
\label{Hfo}
\end{align}
describes two dipole-coupled QDs, where the fermionic operator $d^\dagger_j$ creates an exciton in QD $j$ with the corresponding real frequency $\Omega_j$ ($\hbar=1$ and angular frequencies are used throughout this paper for brevity), and
$g_{\rm F}$ is the strength of the F\"orster coupling between the dots. The second part,
\begin{align}
    H_{\rm IB}&= H_{\rm ph} + \sum_{j=1,2} d^\dagger_j d_j V_j\,,
\label{Hib}
\end{align}
is a generalization of the independent boson~(IB) model to two fermionic modes interacting with a shared bosonic bath, in which
\begin{align}
    H_{\rm ph}=\sum_{\pmb q} \omega_{q} b^\dagger_{\pmb q} b_{\pmb q}\,, \quad V_j&=\sum_{\pmb q} \lambda_{\pmb q,j}\left( b^\dagger_{-\pmb q}+b_{\pmb q}\right)
\label{Vj}
\end{align}	
are, respectively, the free phonon Hamiltonian and the interaction operator of QD $j$ with the bath,
where  $b_{\pmb q}^\dagger$ is a bosonic operator creating a phonon with momentum ${\pmb q}$ and frequency $\omega_{q}$ (with $q=|\pmb q|$, assuming isotropic phonon dispersion). The coupling of the exciton in QD $j$ to the phonon mode ${\pmb q}$ is given by the matrix element $\lambda_{\pmb q,j}$, having a general symmetry property $ \lambda_{\pmb q,j}^\ast= \lambda_{- \pmb q,j}$.  For the same phonon mode,  the matrix elements of identical QDs differ only by a phase factor,
$\lambda_{\pmb q,2}= \lambda_{\pmb q,1} e^{i \pmb q \cdot \pmb d}$, where $\pmb d$ is the vector from the center of the first dot to the center of the second dot, so $d=|\pmb d|$ is the distance between the QDs. The general form of $\lambda_{\pmb q,j}$ in terms of the envelope wave functions $\Psi_{j}$ of the QD excitons is provided in Appendix~\ref{app:gphonon}.

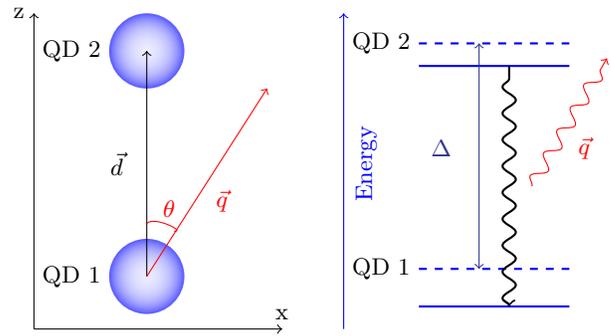
\begin{figure}[t]
\begin{tikzpicture}
\draw[->] (-1.5,-0.7) -- (-1.5,3.5) node at (-1.7,3.5) {z};
\draw[->] (-1.5,-0.7) -- (1.8,-0.7) node at (1.8,-0.5) {x};
\draw [draw=none, fill=blue, fill opacity=0.7] (0,3) circle [radius=0.5] node at (-1,3) [black,fill opacity=1]{QD $2$};
\draw [draw=none, fill=blue, fill opacity=0.7] (0,0) circle [radius=0.5] node at (-1,0) [black,fill opacity=1]{QD $1$};
\foreach \x in {0,...,25} {
\draw [fill=white,fill opacity=0.1,draw=none] (0,3) circle [radius=0.5-0.02*\x];
\draw [fill=white,fill opacity=0.1,draw=none] (0,0) circle [radius=0.5-0.02*\x];
}
\draw[->] (0,0) -- (0,3) node at (-0.4,1.5) {$\vec{d}$};
\draw[->,red] (0,0) -- (1.6,2.5) node at (1,1) {$\vec{q}$};
\draw[red] (0,0.7) to[left=1,in=-0.5] (0.4,0.6) node at (0.3,0.9) {$\theta$};
\end{tikzpicture}
\hspace{0.4cm}
\begin{tikzpicture}
\draw[->,blue] (-2,-0.4) -- (-2,3.8) node at (-1.7,1.8) [rotate=90] {Energy};
\draw node at (-1.5,3.4) {QD $2$};
\draw node at (-1.5,0.4) {QD $1$};
\draw [blue, thick] (-1,3.1) -- (1,3.1);
\draw [blue, thick] (-1,-0.1) -- (1,-0.1);
\draw [blue, thick, dashed] (-1,3.4) -- (1,3.4);
\draw [blue, thick, dashed] (-1,0.4) -- (1,0.4);
\draw[<->, Blue] (-0.2,0.4) -- (-0.2,3.4) node at (-0.7,2) {$\Delta$};
\tikzset{wiggle/.style={decorate, decoration=snake}}
\draw[<-, wiggle, thick] (0.2,-0.1) -- (0.2,3.1);
\draw[->, wiggle, red] (0.5,1.5) -- (1.5,3.2) node at (1.2,2) {$\vec{q}$};
\end{tikzpicture}
\caption{Sketch of investigated system.
Left: Two strongly coupled QDs, which are separated by a distance $|{\bf d}|$, interact with a mode with momentum ${\bf q}$ of the phonon continuum.
Right: A transition (black wiggle) between the polaron-shifted exciton levels (solid blue) of the hybridized QD pair, under the emission of a phonon ${\bf q}$. The detuning $\Delta $ between the  uncoupled exciton levels (dashed blue) 
}
\label{system}
\end{figure}

Knowing the exciton wave functions  $\Psi_{j}(\pmb r_e,\pmb r_h)$, where $\pmb r_e (\pmb r_h)$ is the electron (hole) coordinate relative to the center of the QD, allows us to find the strength of the F\"orster coupling and its dependence on the QD separation $d$,
\be
g_{\rm F}(d)= \int d \pmb r  \int d \pmb r'
\Psi^\ast_1(\pmb r, \pmb r) U_{\rm F} (\pmb r - \pmb r'-\pmb d) \Psi_2(\pmb r', \pmb r')\,,
\label{gF}
\ee
where $U_{\rm F}(\pmb r)$ is the F\"orster potential having the dipole-dipole form 
\begin{align}
U_{\rm F}(\pmb r) = \frac{e^2}{\epsilon_s r^3} \Bigg(|{{\pmb \mu_{cv}}}|^2-3 \frac{|\pmb r \cdot {{\pmb \mu_{cv}}}|^2}{r^2} \Bigg)\,,
\label{mainUF}
\end{align}
$ {{\pmb \mu_{cv}}} = \int_{\mathtt{V}_0} d \pmb r w^*_c(\pmb r) \pmb r w_v(\pmb r)$ is the dipole moment between the Wannier functions  $w_c$ and $w_v$ of the conduction and valence bands, respectively, $r=|\pmb r|$, $\epsilon_s$ is the semiconductor permittivity, and $e$ is the electron charge. The F\"orster potential $U_{\rm F}$ and the coupling parameter $g_{\rm F}$ are derived  microscopically in \App{app:gforster} for the specific case of III-V semiconductor QDs, assuming  symmetry in the $xy$-plane and the QD centers on the $z$-axis, as illustrated in \Fig{system}.

\section{Exact approach}
\label{sec:theory}

The quantum evolution of the system of two F\"orster coupled QDs interacting with the shared bath of acoustic phonons, introduced in \Sec{sec:system}, is described by the master equation
\begin{align}
i\dot{\rho}=&[ H,\rho] + \sum_{j=1,2} i \gamma_j \mathcal{D}[d_j]
\equiv\mathcal{L} \rho,
\label{master}
  \end{align}
in which we have added, using the Lindblad dissipator $\mathcal{D}[d_j](\rho) =2d_j\rho d_j^\dagger-d_j^\dagger d_j\rho-\rho d_j^\dagger d_j$, decay rates $\gamma_j$ of the QD excitons not associated with the phonon coupling, such as radiative decay. The total Liouvillian superoperator $\mathcal{L}$ acting on the density matrix (DM) $\rho(t)$ can be split into two parts, $\mathcal{L}=\mathcal{L}_{\rm F}+\mathcal{L}_{\rm IB}$, where $\mathcal{L}_{\rm IB}$ represents the reversible evolution due to $H_{\rm IB}$, while $\mathcal{L}_{\rm F}$ represents both the reversible evolution due to $H_{\rm F}$ and the irreversible evolution due to $\mathcal{D}$ describing dissipation.

To solve the master equation (\ref{master}), we expand the DM into the basis states of the F\"orster Hamiltonian,
\begin{align}
\rho(t)=\sum_{\eta \xi} \tilde{\rho}_{\eta \xi}(t) |\eta \rangle \langle \xi|\,,
\label{dens-mat-decomp}
\end{align}
where each element $\tilde{\rho}_{\eta \xi}(t)$ includes the phonon degrees of freedom.
The states $|\eta \rangle,|\xi \rangle$ belong to the following reduced basis:
\begin{align}
    |0\rangle, \qquad
    |1\rangle = d_1^\dagger |0\rangle,  \qquad
    |2\rangle = d_2^\dagger |0\rangle,  \qquad
    \label{basis}
\end{align}
where $|0\rangle$ represents the ground state of the QD pair excitonic system.
In the following we use a vector representation of the DM and the corresponding matrix form of \Eq{master} as in~\cite{allcock2022quantum, sirkina2023numeric}. In the present case, the DM is given by a vector $\vec{\rho}(t)$ with components $ \tilde{\rho}_{00},\,\tilde{\rho}_{11},\,\tilde{\rho}_{12},\,\tilde{\rho}_{21},\,\tilde{\rho}_{22}$, denoted by $\rho_i(t)$ with the index $i=0,\,1,\dots4$, in this order.

We assume as initial condition that the system is in it ground state $|0\rangle$ with phonons in thermal equilibrium, and that at $t=0$ QD2 is populated instantaneously, e.g. by excitation via an ultrashort optical pulse. The initial DM is thus given by
$ \rho(0) = |2\rangle \langle 2| \rho_{\rm ph}$, where $\rho_{\rm ph}= e^{-\beta H_\text{ph}}/{\rm Tr}\{e^{-\beta H_\text{ph}}\}$, $\beta=1/(k_B T)$, $T$ is the temperature, $k_B$ is the Boltzmann constant, and the trace is taken over the phonon degrees of freedom. In the vector representation, the initial DM $\vec{\rho}(0)$ then has only one non-zero component $\rho_4(0)=\rho_{\rm ph}$. We focus below on the population of QD1, which is given by
\begin{align}
\mathtt{N}_{12}(t)=  \langle {\rho}_1(t) \rangle_{\rm ph},
\label{N12}
\end{align}
where the expectation value is taken with respect to the phonon bath.
The more general case of the population of QD $j$, assuming that QD $j'$ is initially excited, denoted by $\mathtt{N}_{jj'}(t)$, is considered in Appendix~\ref{app:ecmc}.

\subsection{Trotter's decomposition and linked cluster expansion}
\label{theory:TrotterLCE}

To solve \Eq{master} exactly, we split the time interval between 0 and $t$ into $N$ equal steps of length $\Delta t$. We denote the time at the end of the $n$-th time step by $t_n=n\Delta t$, where $n$ is an integer $0\leqslant n \leqslant N$, so that $\Delta t=t_n-t_{n-1}$, $t_0=0$, and $t_N=t$. We then use Trotter's theorem~\cite{trotter1959product} to separate evolution governed by two non-commuting parts of the Liouvillian superoperator, $\mathcal{L}_{\rm F}$ and $\mathcal{L}_{\rm IB}$, represented by matrices. To implement Trotter's theorem, we treat the evolution of the system during each time step due to each part sequentially,
\be
\vec{\rho}(t_n)= e^{-i \mathcal{L}\Delta t} \vec{\rho}(t_{n-1}) \approx e^{-i \mathcal{L}_{\rm IB}\Delta t} e^{-i \mathcal{L}_{\rm F}\Delta t} \vec{\rho}(t_{n-1})\,.
\label{step-ev}
\ee
so that taking the limit $\Delta t\to0$ provides an exact solution of \Eq{master}. To use the analytical solutions of $H_{\rm F}$ and $H_{\rm IB}$ we introduce a matrix ${M}= e^{-i \mathcal{L}_{\rm F} \Delta t}$, with the explicit form of the F\"orster Liouvillian $\mathcal{L}_{\rm F}$ given by \Eq{LF} of Appendix~\ref{app:ecmc}, and, following Ref.~\cite{sirkina2023numeric}, a vector operator with components ${\Upsilon}_{i_n}$, taking care of the exciton-phonon interaction. A single step in the evolution of the DM then becomes
\begin{align}
{\rho}_{i_{n}}(t_n) \approx &\sum_{i_{n-1}} \tilde{\cal T} \Upsilon_{i_{n}} {M}_{i_{n} i_{n-1}} {\rho}_{i_{n-1}}(t_{n-1})\,,
 \label{evolve-Yvec}
\\
 {\Upsilon}_{i_n}=&\tilde{\cal T} \exp{\left\{ -i \int_{t_{n-1}}^{t_n}  \Tilde{V}_{i_n}(\tau) d\tau \right\}}\,,
  \label{Y-vector}
\end{align}
where $\rho_{i_{n}}$ is $i_n$-th component ($i_n=0,\,1,\dots4$) of the DM vector at time step $n$ 
and $\tilde{\cal T}$ is a specific time ordering operator introduced in~\cite{sirkina2023numeric}, which accounts for the fact that in the quantum evolution, phonon operators act on the DM from both sides (see Appendix~\ref{appendix:trotter} for details). The operator $\Tilde{V}_{i_n}(\tau)$ [defined in \Eq{Vtilde-left-right-vec}] is a linear combination of the exciton-phonon operators $V_j(\tau)$ in the interaction picture [$V_j$ are given in \Eq{Vj}], which takes into account all possibilities of interaction with phonons during the given time interval $t_n-t_{n-1}$, via QD1 ($j=1$) or QD2 ($j=2$), or both.

The $i$-th component of the reduced DM vector at the observation time $t$ then becomes:
\begin{align}
\langle \rho_{i}(t)\rangle_{\rm ph} =&  \sum_{i_{{N-1}}...i_1} {M}_{{i} i_{{N}-1}} ... {M}_{i_2 i_1}   {M}_{i_1 i_0}
\nonumber \\
    &\times \Big\langle
    \tilde{\cal T}{\Upsilon}_{i_{{N}}} ...{\Upsilon}_{i_2} {\Upsilon}_{i_1} 
     \Big\rangle_{\rm ph} ,
\label{P-full-before-LCE-zerotau}
\end{align}
keeping in mind that the initial DM vector has only one non-zero component ${\rho}_{i_0}(0)= \rho_{\rm ph}$. The sum runs over all combinations of intermediate components, each defining a path.
In particular, using $i_0=4$ and $i=1$, \Eq{P-full-before-LCE-zerotau} results in $\mathtt{N}_{12}(t)$, the population of QD1 provided that QD2 was initially excited.
The phonon contribution is encoded in the time-ordered product of the elements ${\Upsilon_{i_n}}$, which is separated from the rest of the expression, so that the linked cluster theorem~\cite{mahan2013many} can be applied in order to treat the exciton-phonon interaction exactly.
This theorem holds for the specific time ordering $\tilde{\cal T}$, as demonstrated in Ref.~\cite{sirkina2023numeric}. Importantly, within the IB model, only second-order connected diagrams (involving two-phonon operators) contribute to the cumulant since the exciton-phonon interaction is diagonal, i.e. it does not contain any direct phonon-assisted transitions within one QD or between the QDs. However, as we discuss in \Sec{sec:analytics} below, phonon-assisted transitions can occur between the hybridized QD states. These transitions are essential for F\"orster transfer and are taken into account exactly in the present approach. Considering the above discussion, we proceed by  writing
\begin{align}
\langle \tilde{\cal T} {\Upsilon}_{i_N}...{\Upsilon}_{i_1} \rangle_{\rm ph} =\exp{\left(\sum_{m=1}^N \sum_{n=1}^N \mathcal{K}_{{i}_{m} {i}_{n}}(|m-n|)\right)},
\label{linked-cluster-expansion}
\end{align}
where
\begin{align}
  \mathcal{K}_{ii'}(s) =-\frac{1}{2} \int_{s\Delta t}^{(s+1)\Delta t} \hspace*{-18pt}d\tau_1 \int_{0}^{\Delta t} \hspace*{-12pt} d\tau_2  \left\langle \Tilde{\cal T} \Tilde{V}_{i}(\tau_1) \Tilde{V}_{i'}(\tau_2) \right\rangle,
    \label{cumulant-Vtilde}
\end{align}
is the cumulant that contains only second-order terms and describes two-time correlations between any pair of time steps, connecting the path segments $n$ and $m$ in \Eq{linked-cluster-expansion}. The cumulant is expressed in \App{appendix:cumulantderivation} as a linear combination [see \Eqsss{cumulants}{C0}{Cs}] of the values of the cumulant functions 
\bea
C_{jj'}(t)&=&-\dfrac{1}{2} \int_{0}^{t} d\tau_1 \int_0^{t} d\tau_2 \left\langle {\cal T} V_{j}(\tau_1) V_{j'}(\tau_2) \right\rangle
\nonumber\\
&=&{I}_{jj'}(t)-i\Omega_{p,jj'}t-S_{jj'}
\label{diag-cumulant-element-fn0}
\eea
on the time grid, where ${\cal T}$ is the standard time-ordering operator. Owing to the finite memory time of the interaction with the phonon continuum, the cumulant functions are separated in the second line of \Eq{diag-cumulant-element-fn0} into the quickly decaying parts,
\be
{I}_{jj'}(t)= \int_0^\infty d\omega \frac{J_{jj'}(\omega)}{\omega^2} \left[ 2\mathcal{N}_\omega\cos(\omega t)+ e^{-i\omega t}\right]\,,
\ee
and the long-time asymptotics described by the polaron shifts
\be
\Omega_{p,jj'}= -\int_0^\infty d\omega \frac{J_{jj'}(\omega)}{\omega}
\label{polaronjj}
\ee
and Huang-Rhys factors
\be
S_{jj'} =\int_0^\infty d\omega \frac{J_{jj'}(\omega)}{\omega^2} ( 2\mathcal{N}_\omega +1
)\,,
\label{HRjj}
\ee
where $J_{jj'}(\omega)=\sum_{\bf q} \lambda_{{\bf q},j}\lambda^\ast_{{\bf q},j'} \delta(\omega-\omega_q)$ are the phonon spectral densities, $\delta(\omega)$ is the Dirac delta function, and $\mathcal{N}_\omega=1/(e^{\beta \omega}-1)$ is the Bose distribution function.

For spherical QDs and a factorized Gaussian model of the exciton wave function with the confinement length $l_j$ in QD $j$ (assuming it is equal for both electron and hole), the phonon spectral densities (see \App{app:gphonon}) take the analytical form
\be
J_{jj'}(\omega)={\cal J}
\omega^3
\exp\left(-\frac{\omega^2}{\omega_{jj'}^2}\right)
\times
\begin{cases}
    1\, \mbox{if}\,  j=j', \mbox{else} \\
    {\rm sinc}(\omega d/v_s) \,, \\
    \end{cases}
    \label{Jjj}
\ee
where ${\cal J}=(D_c-D_v)^2/(4\pi^2\rho_m v_s^5)$,  $D_c$ $(D_v)$ is the deformation potential of the conduction (valence) band,  $\rho_m$ is the material density, $v_s$ is the speed of sound in the material (a linear acoustic phonon dispersion $\omega_q=v_s q$ is assumed),  and $\omega_{jj'}^{-2}={v_s^{-2}}{({l}_j^2+{l}_{j'}^2)/4}$.
The spectral densities \Eq{Jjj} yield the explicit form of the polaron shifts
\be
\Omega_{p,jj}=-{\cal J}\sqrt{\frac{\pi}{2}}\,\left(\frac{v_s}{l_j}\right)^3.
\ee

\subsection{L-Neighbor approach}
\label{theory:LN}

Equations (\ref{P-full-before-LCE-zerotau}) and (\ref{linked-cluster-expansion}) provide an exact 
semi-analytical solution for the DM of F\"orster-coupled QDs in a shared environment. As we apply this solution to semiconductor QDs coupled to bulk acoustic phonons, we note the superohmic character of the coupling spectral density, see e.g. \Eq{Jjj}, which results in a finite memory time $\tau_{\rm IB}$ of the exciton-phonon coupling
(typically $\tau_{\rm IB}$ is a few ps for single QDs in III-V semiconductors) \cite{morreau2019phonon}.  This allows to limit the number of terms within the double summation in \Eq{linked-cluster-expansion}. For two spatially separated QDs, however, the memory time $\tau_{\rm m}$ is generally longer and depends on their separation, see \App{appendix:memory} for details.

We use the $L$-neighbor approach introduced in Ref.~\cite{morreau2019phonon} leading to a significant computational simplification. In this approach, a memory kernel of size $L+1$ is introduced by the cumulant $\mathcal{K}_{i_{n+s} i_{n}}(s)$ with $0\leqslant s \leqslant L$. Then the quantum evolution of the system is described by a tensor $\mathcal{F}^{(n)}$ of rank $L$ or lower, where $1\leqslant n \leqslant N$, with $n=N$ corresponding to the observation time $t=t_N$, according to the discretization scheme. Tensors $\mathcal{F}^{(n)}$ are generated recursively via
\begin{align}
{\mathcal{F}}^{(n+1)}_{i_{s}...i_{1}}=&\sum_{l=1}^h\mathcal{G}^{(s)}_{i_{s}...i_{1}l} {\mathcal{F}}^{(n)}_{i_{s-1}...i_{1}l}\,,
 \label{prop1}
\end{align}
using the initial value
\begin{align}
\mathcal{F}_{i_{L}...i_{1}}^{(1)}=&\sum_{l=1}^h {M}_{i_1 l} \langle {\rho}_{l}(0) \rangle_{\rm ph}\,,
\label{prop0}
\end{align}
where $\rho_{l}(0)$ is the DM vector at $t=0$. Within \Eq{prop1}, $h$ is the number of elements of the DM we need to consider (in the present case $h=5$) and $\mathcal{G}^{(s)}$ is a tensor of rank up to $L+1$, defined recursively by
\begin{align}
\label{G-tensor1}
\mathcal{G}^{(1)}_{i_{1}l}=& {M}_{i_{1} l} e^{\mathcal{K}_{ll}(0) +2\mathcal{K}_{i_1 l}(1)}\,, \\
\mathcal{G}^{(s)}_{i_{s}i_{s-1}...i_{1}l}=& \mathcal{G}^{(s-1)}_{i_{s-1}...i_{1}l} e^{2\mathcal{K}_{i_s l}(s)}\,.
 \label{G-tensor2}
\end{align}
For the time steps up to $n=N-L$, tensors $\mathcal{F}^{(n)}$ and $\mathcal{G}^{(s)}$ have rank $L$ and $L+1$, respectively, with $s=L$ in \Eq{prop1}, but closer to the observation time, their ranks gradually decrease, with $s=N-n$ in \Eqs{prop1}{G-tensor2}.  The reduced DM vector at the observation time $t$ is given by
\begin{align}
\langle {\rho}_{i}(t) \rangle_{\rm ph}=&e^{\mathcal{K}_{i i}(0)} \mathcal{F}^{(N)}_{i}\,,
 \label{propfin}
\end{align}
as follows from \Eqs{P-full-before-LCE-zerotau}{linked-cluster-expansion}.

\section{Analytical model of the F\"orster transfer}
\label{sec:analytics}

In this section we develop an analytical model approximating the F\"orster transfer, combining and enhancing approaches available in the literature, and including a recently derived polaron correction  \cite{morreau2019phonon, sirkina2023numeric, sirkina2023analytic}.
The results of this model are compared in \Sec{sec:results} with the results of the exact treatment presented in \Sec{sec:theory}.

The model applies FGR to phonon-assisted transitions between the hybridized states of the F\"orster-coupled QDs extracted from the long-time asymptotics~\cite{morreau2019phonon,sirkina2023analytic}
of the nearest-neighbor approximation (i.e. $L=1$). This results in the master equation
\be
i \dot{\rho}=
[\tilde{H},\rho] + i \sum_{j=1,2}\gamma_j \mathcal{D}[d_j]
+  i \Gamma_+ \mathcal{D}[p_-^\dagger p_+]
+  i \Gamma_- \mathcal{D}[p_+^\dagger p_-]
\label{master2}
\ee
for a reduced DM $\rho$, simplifying the operators $\tilde{\rho}_{\eta \xi}(t)$ in \Eq{dens-mat-decomp} to functions $\rho_{\eta \xi}(t)$, reflecting the neglect of any phonon bath memory beyond the polaron formation.  Here
\be
\tilde{H}= \tilde{g}\left( d^{\dagger}_1 d_2 + d^{\dagger}_2 d_1\right) + \sum_{j=1,2} \tilde{\Omega}_j d^\dagger_j d_j
\label{H-eff}
\ee
is the polaron-corrected Hamiltonian of the F\"orster-coupled QDs [cf. \Eq{Hfo}], where
\be
\tilde{\Omega}_j={\Omega}_j+ \Omega_{p,jj}\,, \quad \tilde{g}=g_{\rm F} e^{-S/2}\,,
\label{HRg}
\ee
and $S=S_{11}+S_{22}-S_{12}-S_{21}$, with the polaron shifts $\Omega_{p,jj}$ and Huang-Rhys factors $S_{jj'}$ given by \Eqs{polaronjj}{HRjj}. The fermionic operators
\be
p_\pm^\dagger = D_\mp d^\dagger_1 \pm  D_\pm d^\dagger_2\,,
\ee
describing exciton creation in the hybridized states
\be
|\pm\rangle= D_\mp |1\rangle \pm D_\pm |2\rangle\,,
\label{pmstates}
\ee
diagonalize the Hamiltonian \Eq{H-eff},
\be
\tilde{H}=\Omega_+ p_+^\dagger p_+ + \Omega_- p_-^\dagger p_-\,.
\ee
Here $\Omega_{\pm}= (\tilde{\Omega}_1 + \tilde{\Omega}_2 \pm R)/2$ are the energies of the hybrid states $|\pm\rangle$ and $D_\pm = \sqrt{(1 \pm \Delta/R)/2}$, with $\Delta=\tilde{\Omega}_2-\tilde{\Omega}_1$ and  $R= \sqrt{\Delta^2 + 4\tilde{g}^2}$ being, respectively, the detuning \cite{footnote1} 
and Rabi splitting. As demonstrated in \cite{morreau2019phonon,HallPRB25}, the linear exciton-phonon coupling in \Eq{Hib} results in phonon-assisted transitions between the hybrid QD states $|\pm\rangle$, which are represented by the last two Lindblad dissipators in \Eq{master2}, where
\be
\Gamma_+ =  ({\cal N}_{R}+1)  \Gamma_\text{ph} \rmand  \Gamma_- =  {\cal N}_{R}\, \Gamma_\text{ph}
\ee
are the rates of transitions, respectively, from state $|+\rangle $ to state $|-\rangle$ and vice versa, calculated via FGR. For the spectral densities \Eq{Jjj}, $\Gamma_\text{ph}$ takes the explicit form~\cite{HallPRB25} 
\be
\Gamma_\text{ph} =2 \pi {\cal J} D_+^2 D_-^2 R^3
\exp\left(-\frac{l^2 R^2}{2 v_s^2}\right)
[ 1-{\rm sinc}(Rd / v_s]\,.
\label{FGR}
\ee

The master equation (\ref{master2}) describes the population dynamics of the QD pair provided that one of the QDs is initially excited. For example, if the second QD is excited, i.e.  $\rho(0)=|2 \rangle \langle 2|$, the population of the first dot, $\mathtt{N}_{12}(t)$ [see \Eq{N12}], is given by the $|1 \rangle \langle 1|$ component of $\rho(t)$. In the special case \cite{footnote2} of $\gamma_1=\gamma_2\equiv\gamma$ ,
\Eq{master2} has an explicit analytical solution:
\bea
\mathtt{N}_{12}(t) e^{2\gamma t}&=& \frac{1}{2} +\frac{\Gamma_+-\Gamma_-}{\Gamma_++\Gamma_-}\, \frac{\Delta}{2R} -\frac{2\tilde{g}^2}{R^2} e^{-(\Gamma_++\Gamma_-)t}\cos(Rt)
\nonumber\\
&& -\frac{\Delta}{2R} \left(\frac{\Gamma_+-\Gamma_-}{\Gamma_++\Gamma_-}+\frac{\Delta}{R}\right)
e^{-2(\Gamma_++\Gamma_-)t}\,,
\label{N12analyt}
\eea
and other populations are obtained from $\mathtt{N}_{12}(t)$ by using the constant trace of the DM and symmetry arguments. Namely, $\mathtt{N}_{22}(t)=e^{-2\gamma t}-\mathtt{N}_{12}(t)$, while $\mathtt{N}_{21}(t)$ and $\mathtt{N}_{11}(t)$ are obtained from, respectively, $\mathtt{N}_{12}(t)$ and $\mathtt{N}_{22}(t)$ by changing the sign of $\Delta$.

From \Eq{N12analyt} one can obtain, neglecting radiative decay ($\gamma=0$), the value of $\mathtt{N}_{12}(t)$ in the long-time limit,
\be
\mathtt{N}_{12}(\infty)=\frac{1}{2}\left(1 +\frac{e^{\beta R}-1}{e^{\beta R}+1}\, \frac{\Delta}{R}\right)\,,
\label{N12inf}
\ee
quantifying the F\"orster transfer from one QD to the other. This result is derived in \App{app:splitting}, along with a simpler thermal distribution,
\be
\tilde{\mathtt{N}}_{12}(\infty)=\frac{1}{1+e^{-\beta \Delta}}\,,
\label{N12inf0}
\ee
neglecting the hybridization of the QD states.  Clearly, \Eqs{N12inf}{N12inf0} match in the limits of high temperature ($\beta R\ll 1 $), weak F\"orster coupling ($|g_{\rm F}/\Delta|\ll 1$), or zero detuning. These expressions are compared in \Sec{results:dynamics} to the exact population dynamics.

To compare the analytical solution \Eq{N12analyt} with the exact $\mathtt{N}_{12}(t)$ calculated by the Trotter's decomposition approach presented in \Sec{sec:theory}, we fit the exact solution for $\gamma=0$ with the function
\begin{align}
	\mathtt{N}^{\rm fit}_{12}(t) = a + b e^{-t/T_1} - c e^{-t/T_2}\cos{(rt+\Phi)}\,,
	\label{Nfit}
\end{align}
having a form similar to \Eq{N12analyt}, with $a$, $b$, $c$, $T_1$, $T_2$, $r$ and $\Phi$ being fit parameters. We compare the fitted parameters with their analogues in \Eq{N12analyt}, as detailed in \App{app:fit}. Notably, if the analytical result \Eq{N12analyt} was correct, one would expect $\Gamma_++\Gamma_-=1/(2T_{1})=1/T_{2}$, which is not always the case as shown in \Sec{results:sep}, demonstrating a non-vanishing pure dephasing rate $1/T_2^\ast$.

\begin{table}
	\begin{tabular}{ l | c }
		\hline
		Mass density \cite{MuljarovPRL04}& $\rho_m=5.65$ $\mathrm{cm}^{-3}$\,g  \\ 
		Speed of sound \cite{MuljarovPRL04} & $v_s=4.6$\,km $\mathrm{s}^{-1}$  \\ 
		Deformation potential difference \cite{pfeffer1984theory} & $D_c-D_v=-8.0$\,eV  \\
		Exciton confinement radius & $l_{1,2}=l=2.3$\,nm  \\
		Exciton dephasing rate & $\gamma_{1,2}=0$ \\
		Static dielectric constant \cite{LandoltBornstein2001a} & $\epsilon_s=12.53$ \\ 
		Dipole moment \cite{govorov2003spin,govorov2005spin,nazir2005anticrossings} & ${{\mu}}=|{{\pmb \mu_{cv}}}|=0.6$\,nm \\
		\hline
	\end{tabular}
	\caption{
		Parameters used in calculations. These are typical for InGaAs QDs.
	}
	\label{tabparam}
\end{table}

\begin{figure}[b]
	\includegraphics[width=\columnwidth]{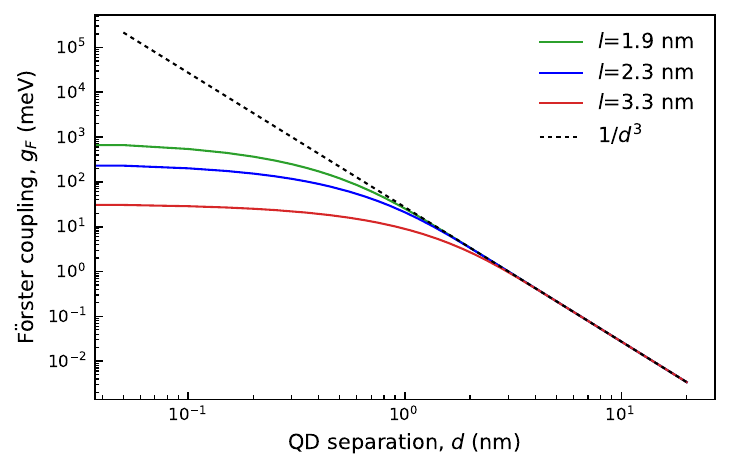}	
	\caption{
		F\"orster coupling strength $g_{\rm F}$ as a function of the QD separation $d$, for different QD sizes: $l=1.9$\,nm (green), $l=2.3$\,nm (blue), and $l=3.3$\,nm (red).  At large $d$ the strength is fitted by $u/d^3$, with $u=27.2$\,meV\,nm$^3$ (black dashed line). 
	}
	\label{fig:coupling}
\end{figure}

\begin{figure*}
	\includegraphics[width=\textwidth]{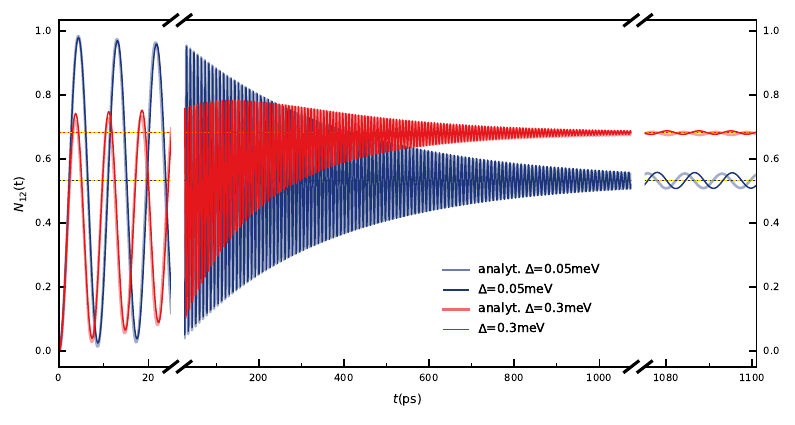}
	\caption[]{Population dynamics $\mathtt{N}_{12}(t)$ of QD1 after QD2 was excited with $\Delta=0.05$\,meV (blue) and 0.30\,meV (red), showing analytical results (thick faint lines) and the exact numerical solution (thin strong lines). Here $d=4.7$\,nm and $T=4$\,K, with other parameters given in \Tab{tabparam}.
		The yellow lines with blue (red) dots show the long-time values $\mathtt{N}_{12}(\infty)$ of the population corresponding to smaller (larger) detuning, extracted from the fit of the exact solution. 
	}
	\label{fig:Nt-agree}
\end{figure*}

\section{Results}
\label{sec:results}

In this section, we focus on the population dynamics $\mathtt{N}_{12}$ of QD1 after QD2 was excited. We compare, for different parameter regimes, the exact solution presented in \Sec{sec:theory} with the analytical approximation developed in \Sec{sec:analytics}. For the exact solution we take numerical results of the $L$-neighbor approach with $L$ up to 16, using the optimization developed in Ref.~\cite{HallArXiv25}. A discussion of the convergence of the numerical results is presented in \App{appendix:memory}.
The default parameters used in the calculations are listed in Table~\ref{tabparam}. Calculations using different values of the exciton confinement radius $l$ and $D_c-D_v$ are presented in \Fig{fig:coupling} and \App{app:sphere} and \ref{app:comparison}.

Of key importance for this work is the realistic F\"orster coupling between QDs given by \Eq{gF} and  calculated microscopically in \App{app:gforster}.  Figure~\ref{fig:coupling} shows results of this microscopic calculation for InGaAs QDs versus their separation $d$, for different QD sizes $l$.
At $d>l$ the F\"orster coupling tends towards the $1/d^3$ dependence of dipole-dipole interaction, as expected for the QD sizes much smaller than $d$, where the QDs behave like point dipoles.
The dipole strength is independent of the QD size for larger distances $d$ in this model of uncorrelated electrons and holes, which is the ``strong confinement''  limit where confinement is much stronger than exciton binding energy. Furthermore, using identical electron and hole wave functions, which have a unity overlap, results in the same dipole moment for different dot sizes at larger $d$.
At shorter distances, there is a reduction of the F\"orster coupling compared to the $1/d^3$, owing to the spatial distribution of the dipoles over the QD size. We note, however, that for such distances, the Coulomb (monopole) interactions between the individual carriers should be taken into account, as well as the electronic tunnel coupling, which is beyond the scope of the present work.

\subsection{Population dynamics: F\"orster transfer}
\label{results:dynamics}

Figure \ref{fig:Nt-agree} shows temporal dynamics of the population $\mathtt{N}_{12}(t)$ for $T=4$\,K, $d=4.7$\,nm and two different values of the detuning, $\Delta =0.05$ and 0.30\,meV, 
for which the analytical and numerical results agree quite well. Note that in these parameters $k_B T=0.34$\,meV is comparable to $g_F=0.26$\,meV, and the nominal Rabi splitting 0.52\,meV is larger than the detuning $\Delta$. The transfer of population occurs due to real phonon-assisted transitions between hybridized QD states and the resulting long-time value depends on the ratio between detuning and temperature (see \App{app:splitting}), as expected from the thermal distributions \Eqs{N12inf}{N12inf0}.
Beatings in $\mathtt{N}_{12}(t)$ occur due to the finite detuning and dipolar coupling between the dots leading to hybridization of their states, as discussed in \Sec{sec:analytics}, and reflect a reversible transfer of population between the QDs. Due to phonons these oscillations are damped, resulting in dephasing and irreversible transfer of population. The long-time values $\mathtt{N}_{12}(\infty)$ provide the fraction of population transferred from QD2 to QD1; they quantify the F\"orster transfer, along with the timescale $T_1$ of the irreversible transfer.  At long times, the populations in \Fig{fig:Nt-agree} almost follow the uncoupled thermal distribution \Eq{N12inf0}, with small deviations increasing at small $d$ or small $T$ and intermediate detunings, see \App{app:splitting}).

The loss of the phase information is crucial for the irreversible transfer of excitation: Without such a loss, there would be a reversible back and forth coherent transfer between the QDs due to their dipolar  coupling. This loss manifests itself as a damping of the oscillations seen in \Fig{fig:Nt-agree}. As a result of exciton-phonon interaction, the phase information is irreversibly lost from the QD pair into the continuum of phonon modes. The exact quantum dynamics, however, shows a more complex non-Markovian process, as discussed in more detail in \Sec{results:sep}, with the phase information being also indirectly exchanged between the QDs, owing to the coherence of the common phonon bath.

In the case presented in \Fig{fig:Nt-agree}, the coupling to phonons is weak, and the quantum dynamics is mostly Markovian, as confirmed by the similarity between the exact results and the analytical approximation, the latter describing a purely Markovian process. A case of even closer agreement is shown in \App{app:sphere}. Considering a wide range of cases investigated, we found mostly Markovian dynamics for lower temperatures, weaker $|D_c-D_v|$ and smaller QD separations, as well as for detunings $\Delta$ similar to or larger than the coupling $g_{\rm F}$. Notably, the oscillations in the analytical result accumulate over time a phase shift relative to the exact results which depends on $\Delta$, as clearly seen in \Fig{fig:Nt-agree} at long times. In what follows, we will explore parameter regimes where the analytical result \Eq{N12inf} no longer provides a good approximation.

\begin{figure}
\raisebox{3.9cm}{a)}\hspace*{-10pt}\includegraphics[width=\columnwidth]{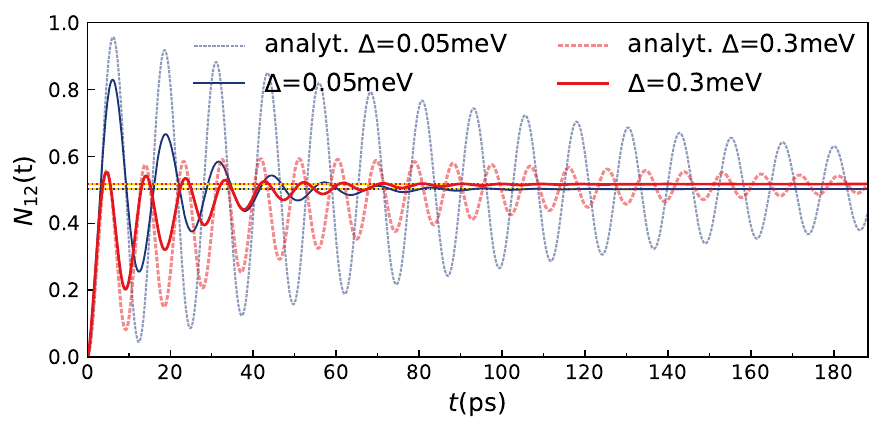}
\raisebox{3.9cm}{b)}\hspace*{-10pt}
\includegraphics[width=\columnwidth]{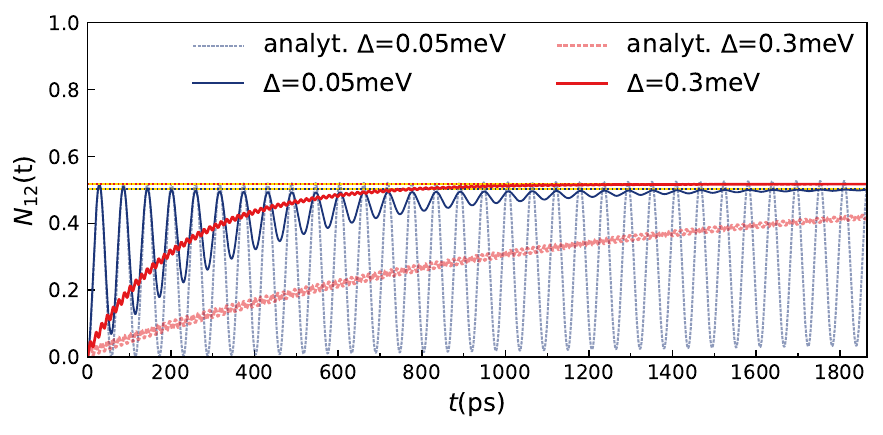}
\caption[]{As \Fig{fig:Nt-agree} but for $T=50$\,K and (a) $d=4.7$\,nm and (b) $d=8.0$\,nm. 
}
\label{fig:Nt-diff}
\end{figure}

\begin{figure}
	\includegraphics[width=\columnwidth]{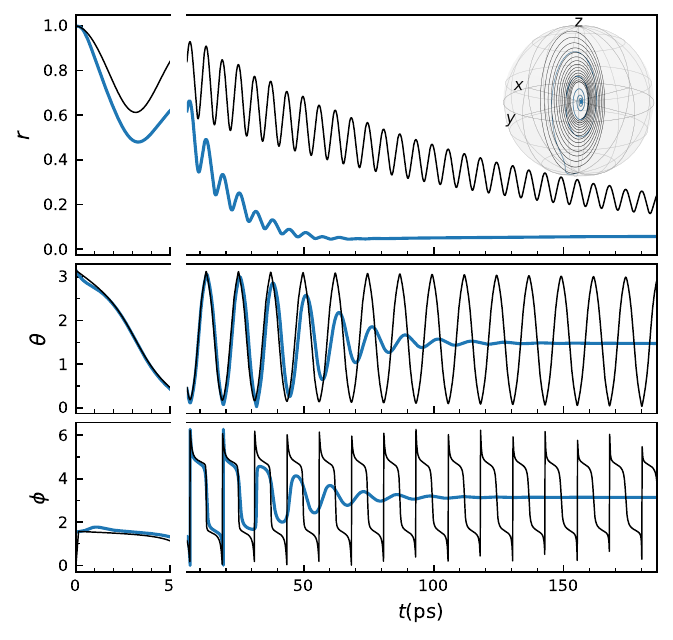}
	\caption[]{Temporal dynamics, showing a comparison of the numerically exact (blue) and analytical (black) results, for the $\Delta=0.05$\,meV case in \Fig{fig:Nt-diff}(a), visualized in the spherical coordinates $(r,\theta,\phi)$ of the Bloch sphere and on the Bloch sphere itself (see inset), for $T=50$\,K, $d=4.7$\,nm. 
	}
	\label{fig:sphere}
\end{figure}

A regime with increased phonon interaction is achieved by increasing the temperature, with examples shown in \Fig{fig:Nt-diff} for $T=50$\,K. Here the analytical result significantly deviates from the exact quantum dynamics, while the long-time values remain in close agreement.

The observed deviation increases going from $\Delta=0.3$\,meV to $\Delta=0.05$\,meV, as well as for or larger QD separation~[\Fig{fig:Nt-diff}(b)]. The damping of the oscillations and the irreversible transfer is significantly underestimated by the analytic result. This may be due to multi-phonon processes, which are not taken into account by FGR but are of increasing importance at higher temperatures due to the larger phonon occupation numbers, as discussed in detail later in \Sec{results:sep}. The corresponding dynamics of the reduced DM given by \Eq{P-full-before-LCE-zerotau} are shown in \Fig{fig:sphere}, using spherical coordinates $(r,\theta,\phi)$ of the Bloch sphere (see inset); the coordinates are $x=\tilde{\rho}_{12}+\tilde{\rho}_{21}$, $y=i(\tilde{\rho}_{12}-\tilde{\rho}_{21})$, and  $z=\tilde{\rho}_{11}-\tilde{\rho}_{22}$ [see \Eqs{dens-mat-decomp}{DMvec}].
This representation gives detailed information about the dynamics via the reduced DM. It shows the population (diagonal elements of the reduced DM corresponding to the $z$-component) and coherence between two exciton states (off-diagonal elements corresponding to the dynamics in the azimuthal plane) of the two QDs. The exact trajectory sets off from the south pole corresponding to the pure state $|2\rangle \langle2|$. While the analytical result shows only a continuous inward spiraling, the exact results show a different initial dynamics over a few picoseconds, comparable to the phonon memory time of the IB model $\tau_{\rm IB}$ \cite{morreau2019phonon}, and then also spirals inwards, but approaching the equilibrium occupation much faster than the analytical result.  These initial changes in the exact result, not visible in the $z$-component [\Fig{fig:Nt-diff}(a)] but seen in the $\theta$-component (\Fig{fig:sphere}), are due to the polaron formation ({\it i.e.} phonon-dressing by exciton-phonon interaction) in QD2 after the creation of an undressed exciton in QD2.

Additional results regarding the Bloch sphere dynamics for other parameters are shown in \App{app:sphere}, and animations are provided in the supplementary information~\cite{Supplement}.
To further exemplify the validity of the exact results presented, we detail in \App{app:comparison} a comparison with a high-quality approximation~\cite{Machnikowski2008quantum}, in which one- and two-phonon processes were taken into account. A good agreement is found, with remaining differences likely due to third- and higher-order phonon processes.

\subsection{Population and polarization decay times}
\label{results:sep}
To analyze the population dynamics, we have fitted the exact results with the fit function \Eq{Nfit}.  Details of the fit and obtained fit parameters are given in \App{app:fit}. Here we concentrate on the decay timescales $T_1$ and $T_2$. Conventionally, $T_1$ is the timescale associated with the decay of population ($z$-component on a Bloch sphere) towards thermal equilibrium, while $T_2$ is the decay of polarization, indicating the coherence between the eigenstates, and is associated with damping of the oscillating term in \Eq{Nfit}, seen in the population signal, as is clear from the analytical model, see \Eqs{master2}{N12analyt}. However, in the exact results the simple lifetime-limited dephasing relationship $T_2=2T_1$ does not hold, and we introduce a pure dephasing time  $T_2^\ast$ given by
\begin{align}
\frac{1}{T_2^\ast}=\frac{1}{T_2}-\frac{1}{2T_1}\,,
\label{T2star}
\end{align}
which is associated with an additional, independent decay of the polarization, not affecting the population.

\begin{figure}
\raisebox{5.2cm}{a)}\hspace*{-12pt}\includegraphics[width=\columnwidth]{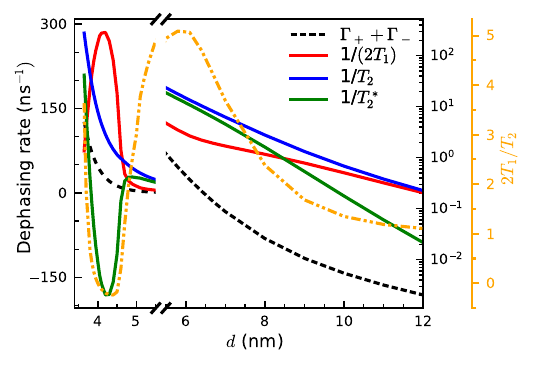}	
\raisebox{4.9cm}{b)}\hspace*{-12pt}\includegraphics[width=\columnwidth]{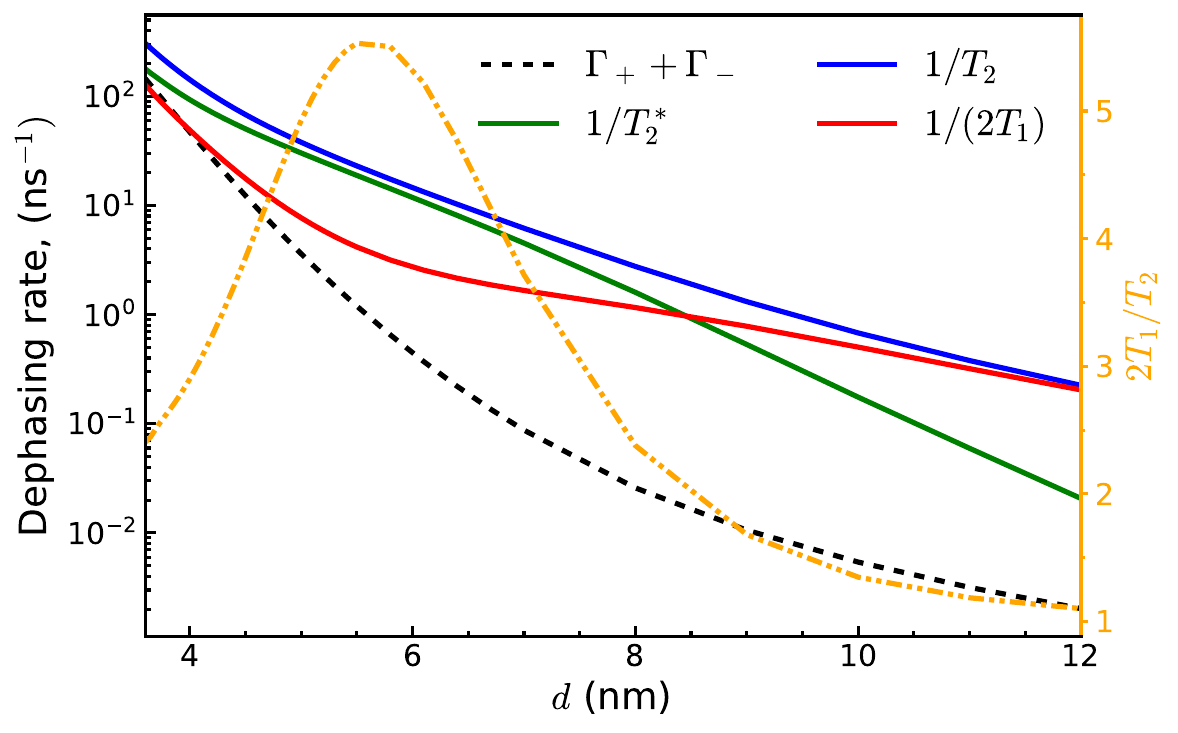}	
\caption[]{Dependence of the decay rates $1/(2T_1)$ (red), $1/T_2$ (blue), and $1/T_2^\ast$ (green) on QD separation $d$, for $T=50$\,K and $\Delta=0.05$\,meV, using the fit \Eq{Nfit} of the exact result, in (a) with $\mathtt{N}_{12}(0)=0$ constraint and in (b) without constraint. Note the change of y scale from linear to logarithmic at the x scale break in (a). The FGR-based rate $\Gamma_++\Gamma_-$ is given as black dashed line. The  $2T_1/T_2$ ratio (orange dash-dotted lines) is also shown, referring to the corresponding second y-scale on the right. 
}
\label{fig:timescales}
\end{figure}

Since we excite QD2 and measure the population of QD1, we have $\mathtt{N}_{12}(0)=0$, which is observed both in the analytical result and the numerically exact data. We therefore require this condition in the fit \Eq{Nfit} that $\mathtt{N}_{12}(0)=0$, providing the constraint $a+b-c\cos(\Phi)=0$ on the fit parameters in \Eq{Nfit}. We fit the exact results as function of QD distance and show in \Fig{fig:timescales}(a) the extracted parameters $1/(2T_1)$ and $1/T_2$, as well as the corresponding pure dephasing rate $1/T_2^\ast$. Other fit parameters are shown in \App{app:fit}. A surprising outcome is a negative pure dephasing rate $1/T_2^\ast$ at small $d$. At $d>2l$, the pure dephasing becomes positive and decays with a power law 
as $d$ is increased further.

To investigate the unexpected negative pure dephasing in \Fig{fig:timescales}(a), we recall that when a system of F\"orster coupled QDs or a QD-cavity systems is excited and measured via the same QD channel, its exciton polarization starts from 1, but the corresponding long-time bi-exponential fit is below 1 at $t=0$, being reduced by the Huang-Rhys factor~\cite{morreau2019phonon,HallPRB25}. This discrepancy  physically corresponds to the polaron formation -- the creation of a phonon cloud around the QD. A similar effect, though not that prominent, may be seen here, since we excite the QD instantaneously, and we have already seen indications of a strongly non-Markovian initial dynamics in the trajectory of the DM on the Bloch sphere. To accommodate this in the fit, we remove the constraint $\mathtt{N}_{12}(0) = 0$ from the fit, and fit only the long-time part of the data, beyond the polaron formation time (thus fitting a different time range). The decay rates extracted from this fit, shown in \Fig{fig:timescales}(b) do not show negative values for $1/T_2^\ast$, and thus provide more physically meaningful results.  We still see a strong deviation of $1/(2T_1)$ from the analytical $\Gamma_++\Gamma_-$ (determined using FGR) almost everywhere except some very short distances. Interestingly, all other parameters of the fit agree surprisingly well with those of the analytical model,  except the phase which is absent in the analytical result, see \Figs{fig:fitphi}{fig:Tfitparams} of \App{app:fit}.
We also note that fitting without constraint the relative errors are reduced by orders of magnitude, see \Fig{fig:tfit} of \App{app:fit} where both fits are illustrated and the errors are shown. This supports that the generalization \Eq{Nfit} of the analytical model \Eq{N12analyt} is suited to describe the population dynamics in the F\"orster transfer.

The pure dephasing rate $1/T_2^\ast$ is specifically prominent around $d=6$\,nm, with the ratio $2T_1/T_2$, shown by the orange curve (right axis) in \Fig{fig:timescales}, exceeding 5, up from its value of 1 in the absence of pure dephasing. This increase is of great importance for the F\"orster transfer mechanism as it implies that a part of the transfer can occur nearly an order of magnitude faster. In fact, the dephasing rate $1/T_2$ determines the damping of the oscillating term in \Eq{Nfit}, expressing the excitation transfer back and forth from one QD to the other (or rather, from one hybrid state to the other). 
Note that owing to the QD state hybridization by the F\"orster coupling, there is one more, even quicker partial transfer, which can be as fast as half the period of Rabi rotations (inversely proportional to the energy splitting), but this is a transfer of a fraction of about $|D_-|^2$ [see \Eq{pmstates}], which small for large detuning. 

The dominance of pure dephasing occurs when the F\"orster coupling $g_{\rm F}$ is about an order of magnitude smaller than the typical energy of local acoustic phonons coupled to the QDs. The  F\"orster coupling value can be seen in \Fig{fig:coupling}, showing $g_{\rm F}=0.13$\,meV at $d=6$\,nm, whereas the typical phonon energy can be estimated from the phonon spectral densities (see \Fig{fig:specdens} in \App{app:spectraldens}) as 1 to 2\,meV. These phonons strongly contribute to real phonon-assisted transitions if the QD energy splitting matches their energy (see \Fig{fig:FGRrates} in \App{app:spectraldens}). However, away from the resonance, such transitions are suppressed, and the dephasing is dominated by virtual phonon-assisted transitions \cite{MuljarovPRL04}. These are involving two or more phonons, which determine the pure dephasing and explain the enhancement in this system with distances around $d=6$\,nm.

\begin{figure}
\includegraphics[width=\columnwidth]{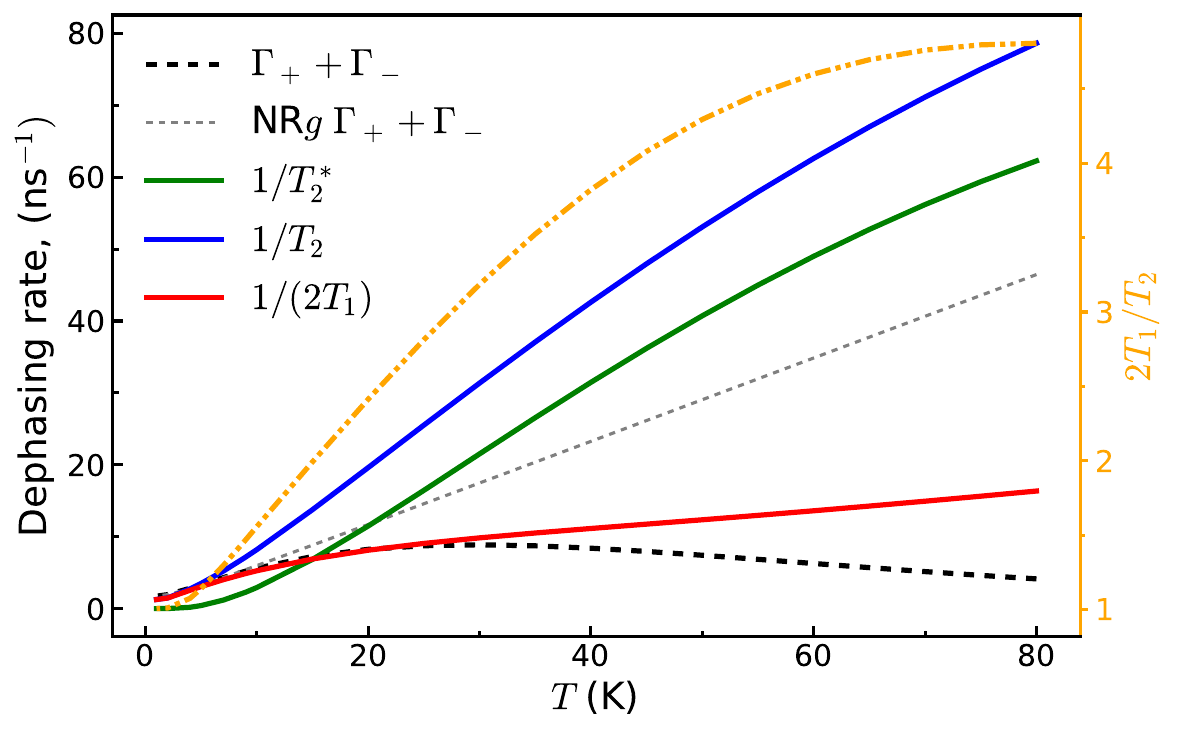}	
\caption{As \Fig{fig:timescales}(b), but versus temperature for $d=4.7$\,nm.
In addition, the gray dashed line shows the FGR result using non-renormalized coupling $g_{\rm F}$ 
}
\label{fig:Ttimescales}
\end{figure}

We now turn to the temperature dependence of the decay rates, for which an example is shown in \Fig{fig:Ttimescales} for a QD distance of $d=$4.7\,nm. Notably, the pure dephasing rate $1/T_2^\ast$ vanishes at zero temperature, which is consistent with the physical picture in which the pure dephasing can only arise as a result of virtual phonon-assisted transitions~\cite{MuljarovPRL04}, and at $T=0$ there are no phonons available. With increasing temperature, the pure dephasing rate becomes significant and dominates the dephasing above $T=15$\,K, as also seen in the ratio $2T_1/T_2$. The lifetime limited dephasing rates $1/(2T_1)$ (red line) and $\Gamma_++\Gamma_-$ (black dotted line) are similar, but do not match exactly because FGR is limited to one-phonon transitions, while the exact solution includes all multi-phonon processes. Interestingly, this leads at low temperatures to $1/T_1$ being lower than $\Gamma_++\Gamma_-$, an effect which was also observed in quantum-dot molecule systems~\cite{Muljarov2005phonon}. Finally, we show in \Fig{fig:Ttimescales} also the non-renormalized (NR) $\Gamma_++\Gamma_-$, which is based on the original F\"orster Hamiltonian \Eq{Hfo}, i.e. without applying the polaron renormalization \Eq{HRg}. Clearly, the NR rate has a poorer agreement with the exact solution at increasing temperatures, which justifies the use of the renormalized F\"orster Hamiltonian \Eq{H-eff} for the analytical model.

\section{Conclusions}
\label{sec:conc}
We have presented an exact path-integral based approach to the quantum dynamics of a pair of spatially separated F\"orster-coupled quantum dots interacting with a continuum of phonon modes. This approach, consisting of Trotter's decomposition and linked-cluster expansion as key elements, provides an exact solution to the F\"orster transfer problem, which was not available before. The F\"orster coupling is taken into account on a rigorous microscopic level, providing a realistic dependence of the coupling constant on the distance between the quantum dots. To analyze the exact solution and to better understand the difference with the existing approaches, we have presented an analytical approximation to the F\"orster transfer dynamics, which combines the best elements of existing theories, including phonon-assisted transitions between hybridized quantum-dot states and polaron corrections. This analysis allowed us to determine an analytical form of the exact population dynamics and to extract the timescales $T_1$, $T_2$, and $T_2^\ast$ of the F\"orster transfer.

We have demonstrated that phonons play a role in the F\"orster transfer mechanism beyond phonon-assisted transitions resulting in a partially reversible process of population transfer from one quantum state to the other. The developed exact solution and the analytical approach allowed us also to reveal a crucial role of pure dephasing in the F\"orster transfer. We found that in the non-perturbative regime of comparable exciton-phonon and F\"orster couplings, the contribution of multi-phonon virtual transitions to the population dynamics dramatically increases with temperature. These virtual transitions are responsible for the pure dephasing which in turn makes the F\"orster transfer nearly an order of magnitude faster in the studied range of low temperatures.

Comparing the exact solution to the analytical model with microscopically determined parameters, we found that all these parameters except the dephasing times $T_1$ and $T_2$
(as well as the phase $\phi$) reasonably well reproduce the exact results.
However, refining the model to include the realistic $T_1$ and $T_2$ resulted in a surprising outcome: The model is either not capable of describing the entire dynamical range of the quantum dot population, or leads to an artifact in a form of a negative pure dephasing time $T_2^\ast$. The main reason for this disagreement is a non-Markovian pure dephasing caused by a polaron formation followed by an optical excitation of the system, which none of the analytical models can cover.

\section*{Acknowledgments}
L.S. acknowledges support from the EPSRC under grant no. EP/R513003/1.
L.H. acknowledges support from the EPSRC under grant no. EP/T517951/1.

\FloatBarrier

\appendix

\section{Microscopic calculation of the F\"orster coupling in QDs}
\label{app:gforster}

Here we provide a microscopic derivation of the F\"orster potential $U_{\rm F}(\pmb R)$ given by \Eq{mainUF} and the distance dependence of the F\"orster coupling constant $g_{\rm F}(d)$ in \Eq{Hfo}. We start from the second-quantized operator of the Coulomb energy in a semiconductor (in Gaussian units),
\begin{align}
\hat{U}=\frac{1}{2} \int d \pmb r \int d \pmb r' \psi^\dagger (\pmb r) \psi^\dagger (\pmb r') \frac{e^2}{\epsilon_s|\pmb r -\pmb r'|} \psi(\pmb r') \psi (\pmb r)\,,
\label{Ugen}
\end{align}
where $\psi (\pmb r)$ is the electron field operator. Expanding the field operator into the Wannier functions of the electron conduction band and doubly-degenerate valence band of the heavy hole denoted by $\sigma=\pm$, we obtain
\begin{align}
\psi (\pmb r)=\sum_{\pmb R} \left[ c_{\pmb R} w_c(\pmb r-\pmb R)+ \sum_\sigma v^\dagger_{{\pmb R},\sigma}w_\sigma(\pmb r- \pmb R) \right],
\label{field}
\end{align}
where $c_{\pmb R}$ and $v^\dagger_{{\pmb R},\sigma}$ are, respectively the electron destruction and hole creation operators, and $\pmb R$ runs over all lattice vectors. In the single-exciton Hilbert space, an exciton state is described by
\begin{align}
\sum_{\pmb R,\pmb R'} \sum_\sigma \Psi_\sigma(\pmb R,\pmb R') c^\dagger_{\pmb R} v^\dagger_{{\pmb R'},\sigma} |0\rangle \,,
\label{ex-state}
\end{align}
where $|0\rangle$ is the electron-hole vacuum state (the state of a fully filled valence band) and $\Psi_\sigma(\pmb R,\pmb R')$ is the first-quantized exciton envelope wave function. Substituting the field operator \Eq{field} into the Coulomb interaction \Eq{Ugen}, and extracting the terms acting on the exciton state \Eq{ex-state} within the same Hilbert space, one can obtain both the direct and exchange Coulomb potentials. Focusing on the exchange term, we obtain the matrix elements of the F\"orster potential which generally mixes different valence-band states,
\be
U^{\sigma\sigma'}_{\rm F} (\pmb R)
= \frac{e^2}{\epsilon_s} \int_{\mathtt{V}_0} d \pmb r \int_{\mathtt{V}_0} d \pmb r' \frac{w^\ast_c(\pmb r)w_\sigma(\pmb r)w_c(\pmb r')w^\ast_{\sigma'}(\pmb r')}{|\pmb r - \pmb r' + \pmb R|},
\label{UF}
\ee
where $\mathtt{V}_0$ is the unit cell volume. To simplify \Eq{UF}, let us consider the multipole expansion,
\begin{align}
\frac{1}{|\pmb R+\pmb r - \pmb r'|} \approx& \frac{1}{R}-\frac{\pmb R \cdot (\pmb r - \pmb r')}{R^3} \nonumber \\
&+\frac{3[\pmb R \cdot (\pmb r - \pmb r')]^2-R^2(\pmb r - \pmb r')^2}{2R^5}
\label{multipole}
\end{align}
neglecting quadrupolar and higher-order terms. Now, owing to the orthogonality of the Wannier functions,
\begin{align}
\int_{\mathtt{V}_0} d \pmb r w^\ast_c(\pmb r) w_\sigma(\pmb r) = 0\,,
\label{FP1}
\end{align}
the only terms in \Eq{multipole} which contribute to the F\"orster potential on this level of approximation are the mixed dipolar terms,
\be
U^{\sigma\sigma'}_{\rm F} (\pmb R)
= \frac{e^2}{\epsilon_s}
\left[-\frac{3 ({\pmb \mu}_\sigma\cdot{\pmb R})({\pmb \mu}^\ast_{\sigma'}\cdot{\pmb R})}{R^5}
+\frac{({\pmb \mu}_\sigma\cdot{\pmb \mu}^\ast_{\sigma'})}{R^3}
\right],
\label{UF1}
\ee
where
\be
{\pmb \mu}_\sigma =\int_{\mathtt{V}_0} d {\pmb r} w^\ast_c(\pmb r) {\pmb r} w_\sigma(\pmb r)\,.
\label{mus}
\ee

In QDs, in the strong confinement limit in $z$-direction, the Wannier functions have the following symmetry:
\begin{align}
w_c(\pmb r) &= \langle \pmb r| s \rangle\,, \\
w_\sigma(\pmb r) &= \frac{1}{\sqrt{2}} \langle \pmb r| x + i\sigma y \rangle\,,
\label{}
\end{align}
which gives a more explicit form of the dipole moment \Eq{mus},
\be
{\pmb \mu}_\sigma=\frac{1}{\sqrt{2}} \langle s| \pmb r| x + i\sigma  y \rangle = \frac{\pmb e_x + i\sigma \pmb e_y}{\sqrt{2}} {{\mu}}\,,
\ee
where by symmetry,
\be
{{\mu}}=|{\pmb \mu}_\sigma|=\langle s|x|x \rangle=\langle s|y|y \rangle
,
\label{dipole}
\ee
and $\pmb e_x$ and $\pmb e_y$ are the unit vectors, respectively, along $x$ and $y$. The F\"orster potential \Eq{FP1} then takes the form
\be
U^{\sigma\sigma'}_{\rm F} (\pmb R)=U_{\rm F} (\pmb R)\delta_{\sigma\sigma'} + U^{\sigma, -\sigma}_{\rm F} (\pmb R)(1-\delta_{\sigma\sigma'})\,,
\label{FPfull}
\ee
where $\delta_{\sigma\sigma'}$ is the Kronecker delta, and the diagonal and off-diagonal elements of the potential are given by
\bea
U_{\rm F} (\pmb R)&=&\frac{e^2 {{\mu}}^2}{\epsilon_s} \frac{Z^2-(X^2+Y^2)/2}{R^5}\,,
\label{FPdiag}\\
U^{\sigma, -\sigma}_{\rm F} (\pmb R)&=&-\frac{e^2 {{\mu}}^2}{\epsilon_s} \frac{3(X+i\sigma Y)^2}{2R^5}\,,
\label{FPoffdiag}
\eea
respectively. The form of the diagonal elements of the potential \Eq{FPdiag}, is identical to \Eq{mainUF} used in the main text, with the dipole moment vector ${\pmb \mu_{cv}}$ of any polarization.

Let us note first of all that, as it is clear from \Eqsss{FPfull}{FPdiag}{FPoffdiag}, in the dipole approximation, the contribution of the Wannier functions to the F\"orster potential is concentrated in a single parameter, the fundamental dipole moment $\mu$, which is well-known in the literature for typical semiconductors. In fact, the actual profile of the Wannier functions is unimportant, only their symmetry matters which we have used in the derivation. Next, we see that the exciton states of different spins (labeled by the index $\sigma$) are coupled by the off-diagonal elements \Eq{FPoffdiag} of the F\"orster potential. However, these off-diagonal elements vanish by symmetry for QDs isotropic in the $xy$ plane and having centers along the $z$-axis, which is the case treated in this work. Let us finally note that the result for the diagonal elements of F\"orster potential  \Eq{FPdiag} is consistent with Refs.~\cite{govorov2003spin,govorov2005spin,nazir2005anticrossings} (see also Refs.~\cite{takagahara1985localization,takagahara2000theory}), but the off-diagonal elements \Eq{FPoffdiag}, which we have also derived, have not been studied in the literature to the best of our knowledge.

To calculate the F\"orster coupling constant $g_{\rm F}(d)$, let us take the wave function of the exciton state in QD $j$ in a factorizable, cylindrically symmetric form,
\begin{align}
\Psi_j (\pmb r_e, \pmb r_h)= \phi_{e,j}(\rho_e) \phi_{h,j}(\rho_h) \psi_{e,j} (z_e) \psi_{h,j} (z_h),
\label{psi0}
\end{align}
where $\pmb r_{e(h)}$ is the electron (hole) coordinate, and we have used the polar coordinates, ${\pmb r}=(\rho,\varphi,z)$.

Assuming that the centers of both QDs lie on the $z$-axis, the off-diagonal elements \Eq{FPoffdiag} vanish by symmetry, and we focus below on the diagonal potential \Eq{FPdiag}, performing in \Eq{gF} the in-plane integration. The auxiliary $z$-dependent F\"orster  potential takes the form
\begin{align}
V_F (z) &= \int d \pmb \rho \int d \pmb \rho' \phi_{e,1} (\rho) \phi_{h,1} (\rho) \nonumber \\
&\times \phi_{e,2} (\rho') \phi_{h,2} (\rho') U_F(\pmb \rho - \pmb \rho', z),
\label{}
\end{align}
which we simplify further by using the Gaussian shape of the in-plane electron and hole wave functions, corresponding to parabolic confining potentials,
\begin{align}
\phi_{e(h),j} (\rho) = \frac{1}{\sqrt{\pi} l_{e(h),j}} e^{-\rho^2/(2l_{e(h),j}^2)}.
\label{phieh}
\end{align}
Now, using for simplicity identical QDs with $l_{e,1}=l_{e,2}=l_e$ and $l_{h,1}=l_{h,2}=l_h$, we obtain
\begin{align}
V_F(z)&=\frac{2 e^2 {{\mu}}^2}{\epsilon_s {\tilde{l}}^3
}  \left( \frac{2 l_e l_h}{l_e^2+l_h^2}    \right)^2 \zeta \left( \frac{|z|}{{\tilde{l}}} \right),
\label{VF}
\end{align}
where
\begin{align}
\zeta(x)&= \Big( x^2+\frac{1}{2} \Big)\sqrt{\pi} e^{x^2} \mathrm{erfc}(x)-x,\\
\mathrm{erfc}(x)&=  \frac{2}{\sqrt{\pi}} \int_x^\infty e^{-t^2} dt,
\end{align}
and ${\tilde{l}}^{-2}=(l_e^{-2}+l_h^{-2})/4$.
Finally, performing the integration along $z$ numerically we find the F\"orster coupling constant
\begin{align}
g_F(d)&= \int_{-\infty}^\infty dz \int_{-\infty}^\infty dz'
 \Big\{ V_F(z-z') \nonumber  \\
&\times \psi_{e,1} (z)  \psi_{h,1} (z)  \psi_{e,2} (z'-d)  \psi_{h,2} (z'-d) \Big\}.
\label{gF2}
\end{align}
In our calculation we used the Gaussian shape of the electron (hole) wave function also in the $z$-direction,
\begin{align}
\psi_{e(h),j} (z) = \left( \frac{1}{\sqrt{\pi} l_{e(h),j}} \right)^{1/2} e^{-z^2/\left(2 l_{e(h),j}^2\right)}\,.
\label{psieh}
\end{align}
Finally, we take for simplicity identical electron and hole confinement radii in each dot $l_{e,j}=l_{h,j}=l_j$ and also use identical dots sizes $l_1=l_2$. We note, however, that this does not have to be the case and the theory presented in this paper is applicable for non-identical, anisotropic QDs and different electron and hole confinement lengths. Furthermore, one can use, if necessary, a more complicated, non-factorizable form of the exciton wave function.

Figures \ref{fig:coupling} and \ref{gF2} show $g_F(d)$ versus distance on a logarithmic and linear scale, respectively.
\begin{figure}[t]
\includegraphics[width=\columnwidth]{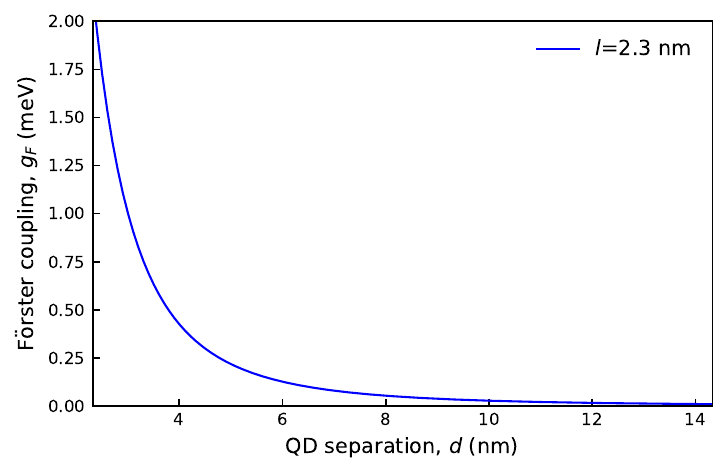}	
\caption{
As \Fig{fig:coupling} but for $l=2.3$\,nm only and shown in linear scale.
}
\label{gF2}
\end{figure}

\section{Calculation of exciton-phonon coupling and spectral densities}
\label{app:gphonon}

Here we calculate the exciton-phonon coupling matrix elements $\lambda_{\pmb q,j}$ and the phonon spectral densities $J_{jj'}(\omega)$ given by \Eq{Jjj} in the Gaussian factorizable model of the exciton wave function in isotropic QDs. We start from the deformation-potential matrix elements in III-V semiconductors~\cite{grosse2008phonons},
\begin{align}
\lambda_{\pmb q,j}=&
\int d \pmb r_e \int d \pmb r_h {\Psi_j}^* (\pmb r_e, \pmb r_h) \nonumber \\
& \times \big\{D_c \pmb \nabla \cdot \pmb u_{\pmb q} (\pmb r_e) - D_v \pmb \nabla \cdot \pmb u_{\pmb q} (\pmb r_h)\big\} \Psi_j (\pmb r_e, \pmb r_h)\,,
\label{lambda1}
  \end{align}
where $\pmb u_{\pmb q}$ is the lattice deformation field associated with the phonon mode $\pmb q$, which we take in the plane-wave form,
\begin{align}
\pmb u_{\pmb q}(\pmb r)=
\pmb e_{\pmb q} \frac{i e^{-i \pmb q \cdot \pmb r} }{\sqrt{2\rho_m v_s \mathtt{V}}}\,,
\label{uq}
  \end{align}
where $\pmb e_{\pmb q}$ is a polarization unit vector (in the direction of $\pmb q$ for LA phonons)
and $\mathtt{V}$ is the sample volume. The matrix element then becomes
\begin{align}
\lambda_{\pmb q,j}=&
\frac{q}{\sqrt{2\rho_m v_s \mathtt{V}}} \int \!d \pmb r
 \big\{D_c |\tilde{\Psi}_{e,j}(\pmb r)|^2 \!-\! D_v |\tilde{\Psi}_{h,j}(\pmb r)|^2 \big\} e^{-i \pmb q \cdot \pmb r}\,,
\label{lambda2}
  \end{align}
where
\be
|\tilde{\Psi}_{e(h),j}(\pmb r_{e(h)})|^2= \int d \pmb r_{h(e)} |\Psi_j (\pmb r_e, \pmb r_h)|^2\,.
\ee

Assuming the factorizable form of the exciton wave function given by \Eq{psi0} with the Gaussian shapes \Eqs{phieh}{psieh}, the exciton-phonon coupling matrix elements take the form
\begin{align}
\label{lambdas1}
\lambda_{\pmb q,1}&=\lambda_0 \sqrt{q} \exp{\{-(q {l}_1/2)^2\}},\\
\lambda_{\pmb q,2}&=\lambda_0 \sqrt{q} \exp{\{-(q {l}_2/2)^2+i \pmb q \cdot \pmb d\}},
\label{lambdas2}
  \end{align}
where
\begin{align}
\lambda_0=\frac{D_c-D_v}{\sqrt{2\rho_m v_s \mathtt{V}}}\,.
\label{constants}
\end{align}
The material parameters used in Eq.(\ref{constants}) are defined in Table~\ref{tabparam}. Note that \Eqs{lambdas1}{lambdas2} were obtained assuming, without loss of generality, that the origin coincides with the center of QD1. 

Using the same exciton model, described in \App{app:gforster} we calculate the phonon spectral densities,
\be
J_{jj'}(\omega)=\sum_{\bf q} \lambda_{{\bf q},j}\lambda^\ast_{{\bf q},j'} \delta(\omega-\omega_q)\,,
\label{J-gen}
\ee
assuming the linear phonon dispersion $q=\omega_q/v_s$.  Converting the discrete sum over ${\bf q}$ into a three-dimensional integral and using spherical coordinates, we obtain
\bea
J_{jj'}(\omega)&=&\frac{\cal J}{2} \omega^3  \exp\left(-\frac{\omega^2}{\omega_{jj'}^2}\right)
\int_0^\pi d\theta \sin\theta
\\
&&\times
\begin{cases}
1 & j=j' \\
\exp(iqd)\cos\theta & j < j' \\
\exp(-iqd)\cos\theta & j > j'\,,
\end{cases}
\label{PhGreensFn-int-0}
\eea
with ${\cal J}$ and $\omega_{jj'}$ defined in \Sec{theory:TrotterLCE}. Performing the integration over the polar angle $\theta$  results in \Eq{Jjj}.

\section{Trotter's decomposition and specific time-ordering}
\label{appendix:trotter}

Here we provide more information on the Trotter decomposition method used to solve \Eq{master}. We consider the evolution according to \Eq{master} over a small time increment $\Delta t$, which is expressed with the help of a specific time-ordering $\tilde{\cal T}$ in a form which separates phonon operators from the rest of the expression, allowing us to apply the cumulant expansion.

We first split the Liouvillian in \Eq{master} into two parts, $\mathcal{L}=\mathcal{L}_{\rm F}+\mathcal{L}_{\rm IB}$, where
\begin{align}
\mathcal{L}_{\rm F} \rho&=[ H_{\rm F},\rho] + \sum_{j=1,2} i \gamma_j \mathcal{D}[d_j]\,,
\label{LiouvilianF}\\
\mathcal{L}_{\rm IB} \rho&=[ H_{\rm IB},\rho]\,.
\label{Liouvilians}
  \end{align}
In the Trotter decomposition, the evolution of the whole system over a small time step $\Delta t$ can be considered independently due to each of these two parts, as described by \Eq{step-ev}. In addition to the matrix ${M}= e^{-i \mathcal{L}_{\rm F} \Delta t}$ already  introduced in \Sec{theory:TrotterLCE}, we consider a diagonal matrix
\bea
W(t_n,t_{n-1}) &=& e^{i H_{\rm ph}t_n} e^{-i H_{\rm IB}(\Delta t)} e^{-i H_{\rm ph}t_{n-1}}\,,
\label{W-matrix}\\
&=& {\cal T} \exp{\left\{ -i \int_{t_{n-1}}^{t_n}  d\tau \sum_{j=1,2} V_{j}(\tau) d^\dagger_j d_j\right\}}
\nonumber
\eea
describing the effect of $\mathcal{L}_{\rm IB}$, where $V_{j}(\tau)= e^{i H_{\rm ph} \tau} V_j e^{-i H_{\rm ph} \tau}$ is  the interaction representation of the exciton-phonon coupling operator $V_j$ defined in \Eq{Vj}.
Then the evolution over a single time step in \Eq{step-ev} can be written for the vector components of the DM as
\bea
{\rho}_{i_n}(t_n)  &\approx& \sum_{i_{n-1}}
W_{i_{n} i_{n}}(t_n,t_{n-1})
{M}_{i_n i_{n-1}} {\rho}_{i_{n-1}}(t_{n-1})
\nonumber\\
&&\times {W}^\dagger_{i_{n} i_{n}} (t_n,t_{n-1})\,,
\label{evolve-component}
\eea
where we have used the fact that $W_{i_{n} i_{n}}(t_n,t_{n-1}) =W^\dagger_{i_{n} i_{n}} (t_{n-1},t_n)$. Using the specific time-ordering operator $\Tilde{\cal T}$ (which ensures that all phonon operators in $W$ [${W}^\dagger$] stand to the left [right] of $\vec{\rho}$ in normal [inverse] order) introduced in \Sec{theory:TrotterLCE}, \Eq{evolve-component} takes the form
\bea
{\rho}_{i_n}  &\approx &\sum_{i_{n-1}} \tilde{\cal T}
W_{i_{n} i_{n}} {W}^\dagger_{i_{n} i_{n}}
  {M}_{i_n i_{n-1}} {\rho}_{i_{n-1}}
\nonumber \\&=&
\sum_{i_{n-1}} \tilde{\cal T} \Upsilon_{i_{n}}  {M}_{i_{n} i_{n-1}} {\rho}_{i_{n-1}}\,,
 \label{evolve-Tordering}
\eea
where the operator $\Upsilon=\tilde{\cal T} W W^\dagger$ is given explicitly by \Eq{Y-vector}. The interaction $\Tilde{V}_{i_n}(\tau)$ within \Eq{Y-vector} takes the form
\be
\Tilde{V}_{i_n}(\tau)=  \alpha_{i_n} V_1(\tau)+\chi_{i_n} V_2(\tau)- \beta_{i_n} V_1(\tau) - \nu_{i_n} V_2(\tau)
\label{Vtilde-left-right-vec}
\ee
for $t_{n-1} \leqslant \tau \leqslant t_{n}$.  Here we have introduced the left vectors $\vec{\alpha}$, $\vec{\chi}$ and right vectors  $\vec{\nu}$, $\vec{\beta}$, which are used to describe the evolution of the DM by a single operator $\Tilde{V}_{i_n}(\tau)$.
These vectors contain information about which system states are coupled to phonons. Their components have Boolean form  and are given explicitly by \Eq{alpha-nu}.
In particular, vectors  $\vec{\alpha}$ ($\vec{\beta}$) and $\vec{\chi}$ ($\vec{\nu}$)
describe coupling to phonons, respectively, via QD1 and QD2, with phonon operators acting on $\vec{\rho}$ from the left (right).

\section{Excitation, measurement, and system Liouvillian}
\label{app:ecmc}
Here we provide explicit expressions for the Liouvillian $\mathcal{L}_{\rm F}$, left and right vectors, $\vec{\alpha}$, $\vec{\chi}$ and $\vec{\beta}$, $\vec{\nu}$, as well as vectors describing the excitation and measurement channels.

The population dynamics is given by the time dependence of diagonal elements of the DM, $\tilde{\rho}_{11}$ and $\tilde{\rho}_{22}$, see \Eq{dens-mat-decomp}. However, owing to the coherence properties of the system, these elements are related to the off-diagonal elements $\tilde{\rho}_{12}$ and $\tilde{\rho}_{21}$. In addition, the ground state population $\tilde{\rho}_{00}$ is accessible via the Lindblad decay. Then, according to \Eqs{Hfo}{LiouvilianF}, the DM vector and the Liouvillian matrix $\mathcal{L}_{\rm F}$ take the form
\be
\vec{\rho}
=\left(
  \begin{array}{cccc}
  \rho_{0} \\
  \rho_{1} \\
  \rho_{2} \\
  \rho_{3} \\
  \rho_{4}\\
  \end{array}
\right)
\equiv
\left(
  \begin{array}{cccc}
  \tilde{\rho}_{00} \\
  \tilde{\rho}_{11} \\
  \tilde{\rho}_{12} \\
  \tilde{\rho}_{21} \\
  \tilde{\rho}_{22}\\
  \end{array}
\right)
\label{DMvec}
\ee
and
\be
\small
 \mathcal{L}_{\rm F}=
\begin{pmatrix}
   {0} & {2i\gamma_1} & {0} & {0} & {2i\gamma_2}\\
   {0} & {-2i\gamma_1} & {-g_{\rm F}} & {g_{\rm F}} & {0}\\
   {0} & {-g_{\rm F}} & {-\Delta-i\gamma_1-i\gamma_2} & {0} & {g_{\rm F}}\\
   {0} & {g_{\rm F}} & {0} & {\Delta-i\gamma_1-i\gamma_2} & {-g_{\rm F}}\\
   {0} & {0} & {g_{\rm F}} & {-g_{\rm F}} & {-2i\gamma_2}\\
\end{pmatrix},
\label{LF}
\ee
respectively, where $\Delta=\Omega_2-\Omega_1$ is the detuning.

Using the basis \Eq{dens-mat-decomp} and the vector representation \Eq{DMvec}, the left and right vectors contributing to the interaction \Eq{Vtilde-left-right-vec} are defined as
\begin{align}
\vec{\alpha}=
\begin{pmatrix}
   {0}\\
   {1}\\
   {1}\\
   {0}\\
   {0}\\
\end{pmatrix},
\quad
\vec{\chi}=
\begin{pmatrix}
   {0}\\
   {0}\\
   {0}\\
   {1}\\
   {1}\\
\end{pmatrix},
\quad
\vec{\beta}=
\begin{pmatrix}
   {0}\\
   {1}\\
   {0}\\
   {1}\\
   {0}\\
\end{pmatrix},
\quad
\vec{\nu}=
\begin{pmatrix}
   {0}\\
   {0}\\
   {1}\\
   {0}\\
   {1}\\
\end{pmatrix}.
\label{alpha-nu}
\end{align}
These vectors are linearly dependent, $\vec{\alpha}-\vec{\beta}=\vec{\nu}- \vec{\chi}$.

We also generalize here the result for the population $\mathtt{N}_{12}(t)$ [given by Eq.(\ref{N12}) and  Eq.(\ref{P-full-before-LCE-zerotau})] to arbitrary excitation and measurement channels. For QD $j'$ excited at time $t=0$, the population of QD $j$ at time $t$ is given by
\begin{align}
\mathtt{N}_{jj'}(t)&=  \sum_{i_{N}} O_{i_N}^{(j)} \sum_{i_{{N-1}}...i_1} {M}_{{i_{{N}}} i_{{N}-1}} ... {M}_{i_2 i_1}  \nonumber \\
&\times \sum_{i_0} {M}_{i_1 i_0} Q_{i_0}^{(j')}
    \Big\langle
    \tilde{\cal T}{\Upsilon}_{i_{{N}}} ...{\Upsilon}_{i_2} {\Upsilon}_{i_1}
     \Big\rangle_{\rm ph} ,
\label{Njjp}
\end{align}
where the components $Q_{i_0}^{(j')}$ of the excitation vector $\vec{Q}^{(j')}$ give the corresponding components of the reduced DM at $t=0$. The observation channel $\vec{O}^{(j)}$ allows us to select specific elements of the DM. The excitation and measurement vectors for QD1 and QD2 are given, respectively, by
\begin{align}
{\vec{Q}}^{(1)}={\vec{O}}^{(1)}=
\begin{pmatrix}
   {0}  \\
	{1}  \\
	{0}  \\
	{0}  \\
	{0}
\end{pmatrix},
\hspace{0.5cm}
{\vec{Q}}^{(2)}={\vec{O}}^{(2)}=
\begin{pmatrix}
   {0}  \\
	{0}  \\
	{0}  \\
	{0}  \\
	{1}
\end{pmatrix}.
\label{Q1-x-c}
\end{align}

The vectors for the measurement channels in the Bloch sphere dynamics, presented in \Figsss{fig:sphere}{fig:bloch-diff}{fig:agree-better-sph}, are
\begin{align}
{\vec{O}^{(x)}}=
\begin{pmatrix}
   {0}  \\
	{0}  \\
	{1}  \\
	{1}  \\
	{0}
\end{pmatrix},
\quad
{\vec{O}^{(y)}}=
\begin{pmatrix}
   {0}  \\
	{0}  \\
	{i}  \\
	{-i}  \\
	{0}
\end{pmatrix},
\quad
{\vec{O}^{(z)}}=
\begin{pmatrix}
   {0}  \\
	{1}  \\
	{0}  \\
	{0}  \\
	{-1}
\end{pmatrix}.
\end{align}

\section{Explicit form of the cumulants}  
\label{appendix:cumulantderivation}
Here we derive an explicit expression for the cumulants ${\mathcal{K}}_{{i}_{m} {i}_{n}}(|m-n|)$ contributing to the linked cluster expansion \Eq{linked-cluster-expansion} in terms of the left and right vector components and the values of the cumulant functions $C_{jj'}(t_n)$ on the time grid. To calculate the cumulants
\be
{\mathcal{K}}_{{i}_{m} {i}_{n}}(|m-n|) =-\frac{1}{2} \int_{t_{m-1}}^{t_m} \!\!\!d\tau_1  \int_{t_{n-1}}^{t_n} \!\!\! d\tau_2  \Big\langle \Tilde{\cal T} \Tilde{V}_{i_m}(\tau_1) \Tilde{V}_{i_n}(\tau_2)\Big\rangle
    \label{cumulant-Vtilde-der}
\ee
we first expand
\bea
&&
\left\langle \Tilde{\cal T} \Tilde{V}_{i_m}(\tau_1) \Tilde{V}_{i_n}(\tau_2)\right\rangle =
\\
&& \left\langle {\cal T}[ \alpha_{i_m} V_1(\tau_1)+\chi_{i_m} V_2(\tau_1)]
[ \alpha_{i_n} V_1(\tau_2)+\chi_{i_n} V_2(\tau_2)] \right\rangle
   \nonumber \\
&&- \left\langle [ \beta_{i_n} V_1(\tau_2)+\nu_{i_n} V_2(\tau_2)]
[ \alpha_{i_m} V_1(\tau_1)+\chi_{i_m} V_2(\tau_1)] \right\rangle
   \nonumber \\
&&- \left\langle [ \beta_{i_m} V_1(\tau_1)+\nu_{i_m} V_2(\tau_1)]
[ \alpha_{i_n} V_1(\tau_2)+\chi_{i_n} V_2(\tau_2)] \right\rangle
   \nonumber \\
&&+ \left\langle {\cal T}_{\rm inv}[ \beta_{i_m} V_1(\tau_1)+\nu_{i_m} V_2(\tau_1)]
[ \beta_{i_n} V_1(\tau_2)+\nu_{i_n} V_2(\tau_2)] \right\rangle
   \nonumber
\eea
using the definitions of the interaction $\Tilde{V}_{i}(\tau)$ given by \Eq{Vtilde-left-right-vec} and the time-ordering operator $\Tilde{\cal T}$, reducing the effect of the latter to the standard ${\cal T}$ and inverse ${\cal T}_{\rm inv}$ time-ordering operators, as well as to a fixed order of operators.
Next, we introduce the phonon propagators,
\bea
D_{jj'}(\tau_1-\tau_2)&=&\left\langle {\cal T} V_j(\tau_1) V_{j'}(\tau_2)\right\rangle
\nonumber\\
&=& \int_0^\infty d\omega \; J_{jj'}(\omega) D(\omega,\tau_1-\tau_2)\,,
\eea
where
\begin{equation}
D(\omega,t) = (N_\omega +1) e^{-i\omega |t| }  + N_\omega e^{i\omega |t|}
\end{equation}
and $J_{jj'}(\omega)$ are the phonon spectral densities defined by \Eq{J-gen}. Owing to the general symmetry properties, $D(\omega,-t)=D(\omega,t)$ and $J_{j'j}(\omega)=J_{jj'}(\omega)$, we find
\be
D_{jj'}(\tau_1-\tau_2)=D_{j'j}(\tau_1-\tau_2)=D_{jj'}(\tau_2-\tau_1)\,.
\ee
Finally, introducing cumulant elements
\be
K_{jj'}(|m-n|) =-\frac{1}{2} \int_{t_{m-1}}^{t_m}  d\tau_1  \int_{t_{n-1}}^{t_n}  d\tau_2  {D}_{jj'}(\tau_1-\tau_2)\,,
\label{cumulant-element-defn-rectangular}
\ee
and noting that
\be
\left\langle {\cal T}_{\rm inv} V_{j}(\tau_1) V_{j'}(\tau_2) \right\rangle= {D}_{{j}{j'}}^\ast (\tau_1-\tau_2)
\ee
and
\be
\left\langle  V_j(\tau_1) V_{j'}(\tau_2) \right\rangle
=
\begin{cases}
{D}_{jj'}(\tau_1-\tau_2) \quad \ \tau_1>\tau_2\,, \\ {D}_{jj'}^\ast (\tau_2-\tau_1) \quad \ \tau_1<\tau_2 \,,  \\
\end{cases}
\label{PhGreensFn}
\ee
allows us to evaluate
\begin{align}
-\frac{1}{2}\int_{t_{m-1}}^{t_m}  d\tau_1  & \int_{t_{n-1}}^{t_n}  d\tau_2   \left\langle  V_j(\tau_1) V_{j'}(\tau_2) \right\rangle = \nonumber \\
    & \begin{cases}
      K_{jj'}(|m-n|)  \ &m>n\,, \\
      K_{jj'}^\ast(|m-n|) \ &m<n\,, \\
      \frac{1}{2}[ K_{jj'}(0)+ K_{jj'}^\ast(0)]  \ &m=n\,.
    \end{cases}
    \label{Ktild}
\end{align}
Then we obtain with the help of \Eq{alpha-nu} an explicit expression for the cumulants:
\bea
{\mathcal{K}}_{{i}_{m} {i}_{n}}(s)&=&
(\alpha_{i_m} -\beta_{i_m}) [\alpha_{i_n} K_{11}(s) -\beta_{i_n} K_{11}^\ast(s)
\nonumber\\
&&
-(\alpha_{i_n} -\chi_{i_n}) K_{12}(s)
+(\beta_{i_n} -\nu_{i_n}) K_{12}^\ast(s)
\nonumber\\
&&
-\chi_{i_n} K_{22}(s) +\nu_{i_n} K_{22}^\ast(s)]\,,
\label{cumulants}
\eea
where $s=m-n\geqslant0$. In the remaining case $m-n<0$, one can use the fact that ${\mathcal{K}}_{{i}_{m} {i}_{n}}(|m-n|)={\mathcal{K}}_{{i}_{n} {i}_{m}}(|m-n|)$, following from the general symmetry of $D(\omega,t)$ and $J_{jj'}(\omega)$.

In practice, we need to calculate only a small number of distinct cumulant elements. In fact, using the cumulant functions $C_{jj'}(t)$ defined by \Eq{diag-cumulant-element-fn0},
we obtain
\be
K_{jj'}(0) = C_{jj'}(\Delta t)\,.
\label{C0}
\ee
The remaining cumulant elements with $0<s\leqslant L$ are then found recursively as
\bea
K_{jj'}(s) &= &\frac{1}{2}\left[C_{jj'}\big((s+1)\Delta t\big) - (s+1)K_{jj'}(0)\right]
\nonumber\\
&&-\sum_{h=1}^{s-1} (s+1 -h) K_{jj'}(h).
\label{Cs}
\eea

\section{Choosing the memory time and convergence parameters}
\label{appendix:memory}

In the $L$-neighbor approach, we use a finite memory time of the form $\tau_{\rm m}=t_D +\zeta \tau_{\rm IB}$ (where $\tau_{\rm IB}= \sqrt{2} \pi l/v_s$ and $1\leq \zeta  \leq 2$), which accounts for a delayed decay of the cumulant function $C_{12}(t)$ defined in \Eq{diag-cumulant-element-fn0}.
We use the delay time as given by $t_D \approx d/v_s$.  The value of $\zeta $ can be chosen by assessing its impact on convergence. We choose to use $\zeta=1.2$ in all simulations.
The number of neighbors $L$, and the associated time step $\Delta t = \tau_{\rm m}/(L+1)$ must be chosen to satisfy $\Delta t \ll \tau_{\rm F}$, where $\tau_{\rm F} \approx \pi / g_{\rm F}$ is the timescale associated with the F\"orster interaction (at zero detuning).

\begin{figure}	
\includegraphics[width=\columnwidth]{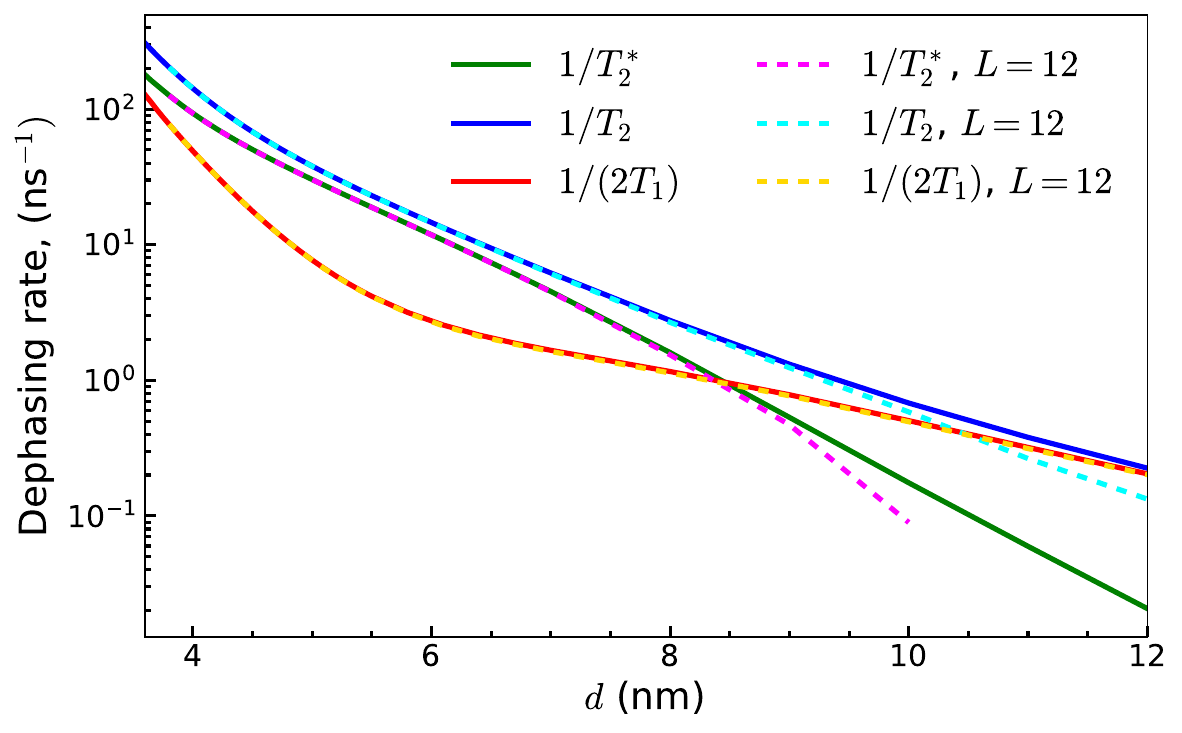}
\caption{
Comparison of $L=16$ optimized numeric results shown in  \Fig{fig:timescales}(b) with $L=12$ unoptimized numeric results.
}
\label{fig:Lcomp}
\end{figure}

To ensure that our numerical results are exact, we have used the optimization~\cite{HallArXiv25} based on a singular value decomposition with low singular value threshold of $10^{-8}$. This allows us to reach $L=16$. We note, however, that the exact results for smaller dot separations $d\leq 8$~nm show no visual difference (see \Fig{fig:Lcomp}) to the unoptimized $L=12$ calculation (which is already well converged), except for the refined splitting energies~$r_{\rm ac}$, which are extracted from the optimized exact results. These are shown in \Fig{fig:energies-T-d} of \App{app:splitting}.
For $d>8$~nm $L=12$ results are not fully converged and underestimate the dephasing rate, although the decay rate is well reproduced.

\section{Spectral densities and FGR rates}
\label{app:spectraldens}

\begin{figure}	
\includegraphics[width=\columnwidth]{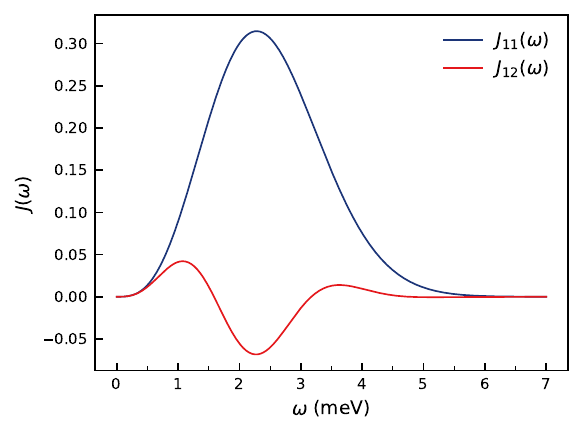}
\caption{
Spectral densities for $d=6$\,nm, $T=50$\,K,  other parameters in Table~\ref{tabparam}.
}
\label{fig:specdens}
\end{figure}

\begin{figure}	
\includegraphics[width=\columnwidth]{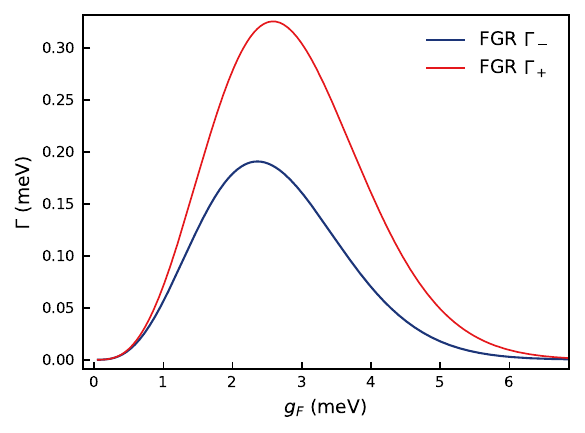}
\caption{
The dependence of FGR transition rates in \Eq{FGR} on $g_F$ using $\Delta=0.05$\,meV and $T=50$\,K.
}
\label{fig:FGRrates}
\end{figure}

This Appendix provides spectral densities for the parameter regime with maximized pure dephasing, discussed in \Sec{results:sep} of the main text. These are shown in \Fig{fig:specdens}. Additionally, the dephasing rates given by  FGR are shown in \Fig{fig:FGRrates} over a range of different F\"orster couplings $g_{\rm F}$, for  $T=50$\,K and $\Delta=0.05$\,meV.

\section{Bloch sphere representation and more results on the F\"orster dynamics}
\label{app:sphere}

\begin{figure}[t]	
\raisebox{3.8cm}{a)}\hspace*{-12pt}
\includegraphics[width=0.48\columnwidth]{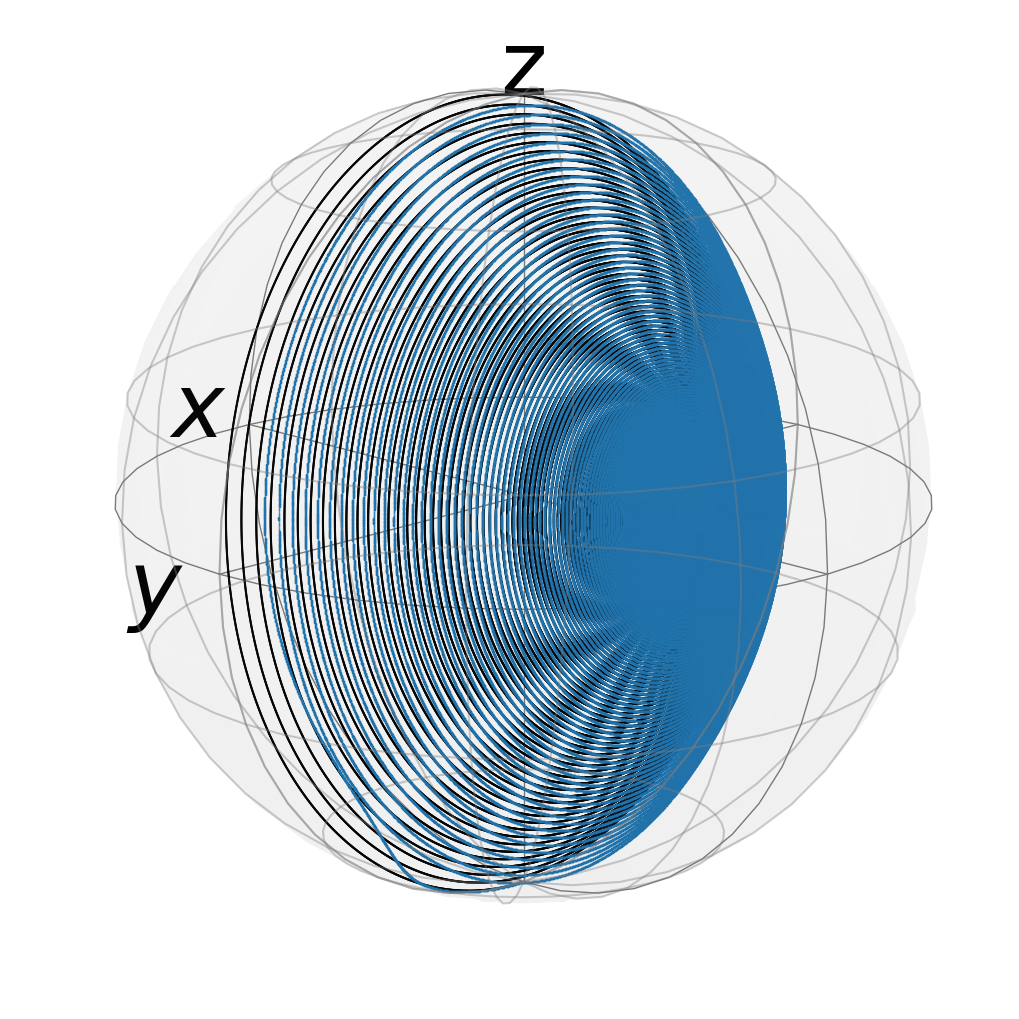}
\raisebox{3.8cm}{b)}\hspace*{-12pt}
\includegraphics[width=0.48\columnwidth]{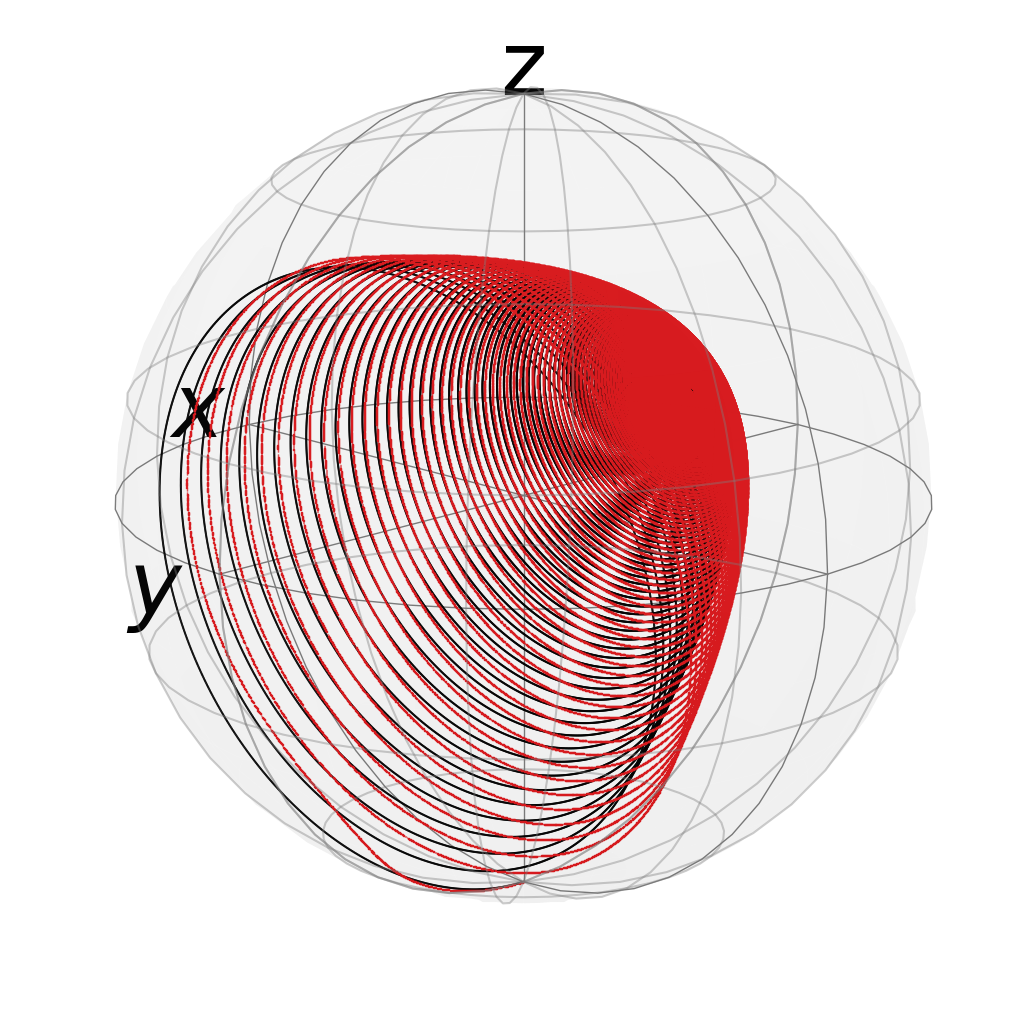}
\raisebox{3.8cm}{c)}\hspace*{-12pt}
\includegraphics[width=0.48\columnwidth]{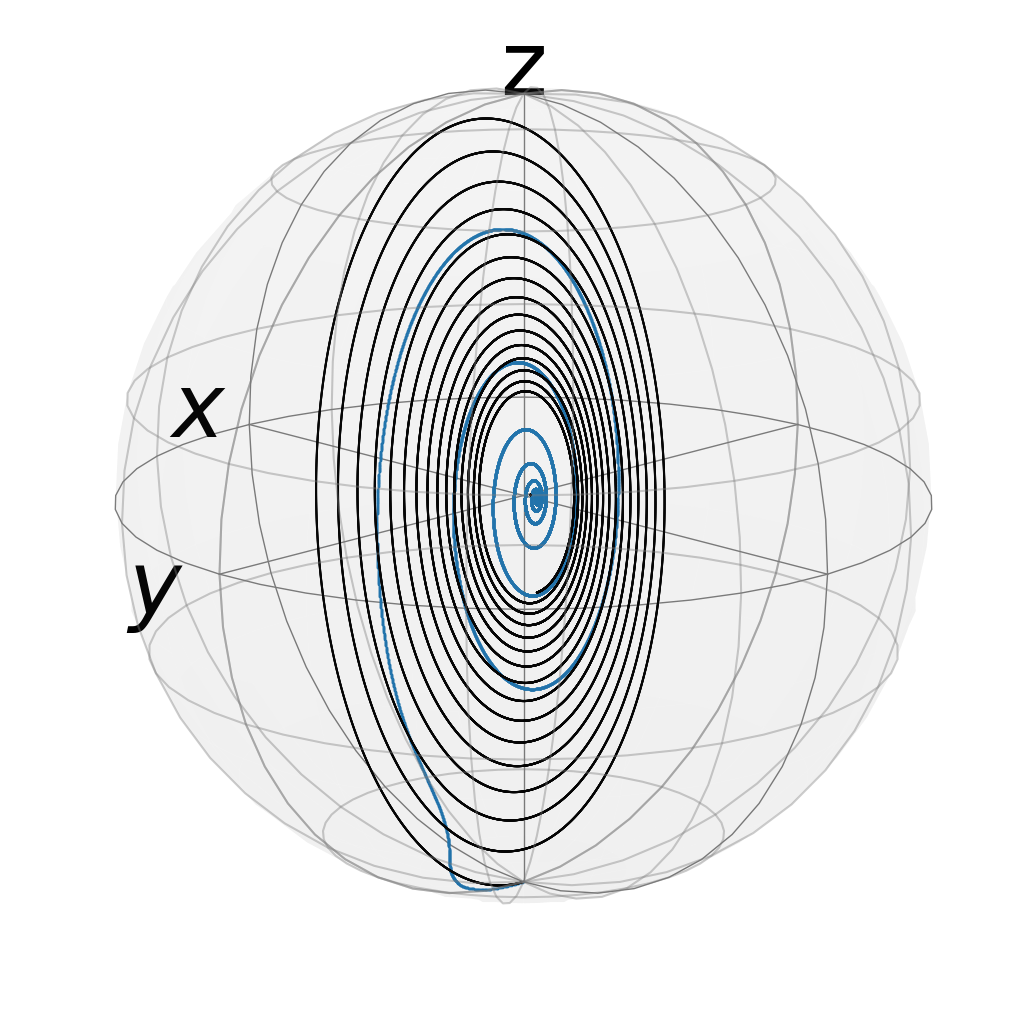}
\raisebox{3.8cm}{d)}\hspace*{-12pt}
\includegraphics[width=0.48\columnwidth]{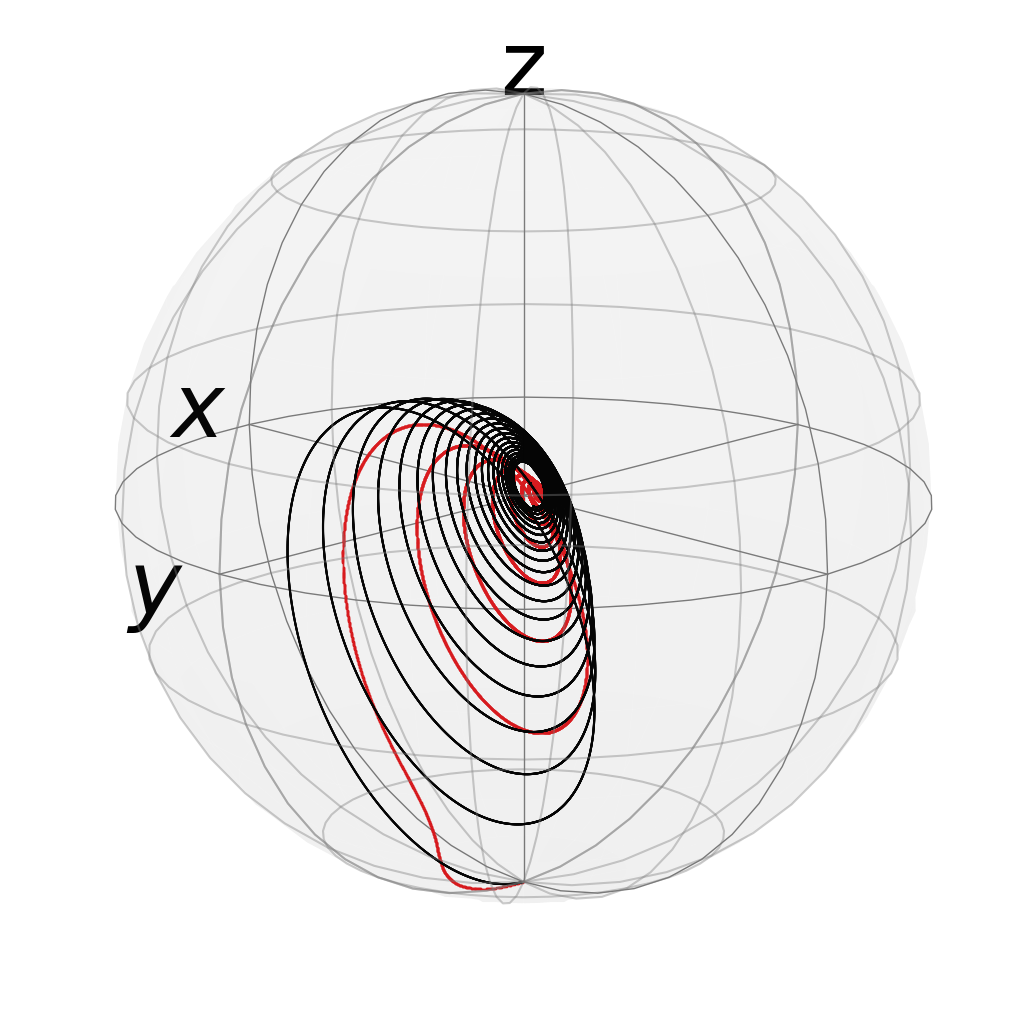}
\raisebox{3.8cm}{e)}\hspace*{-12pt}
\includegraphics[width=0.48\columnwidth]{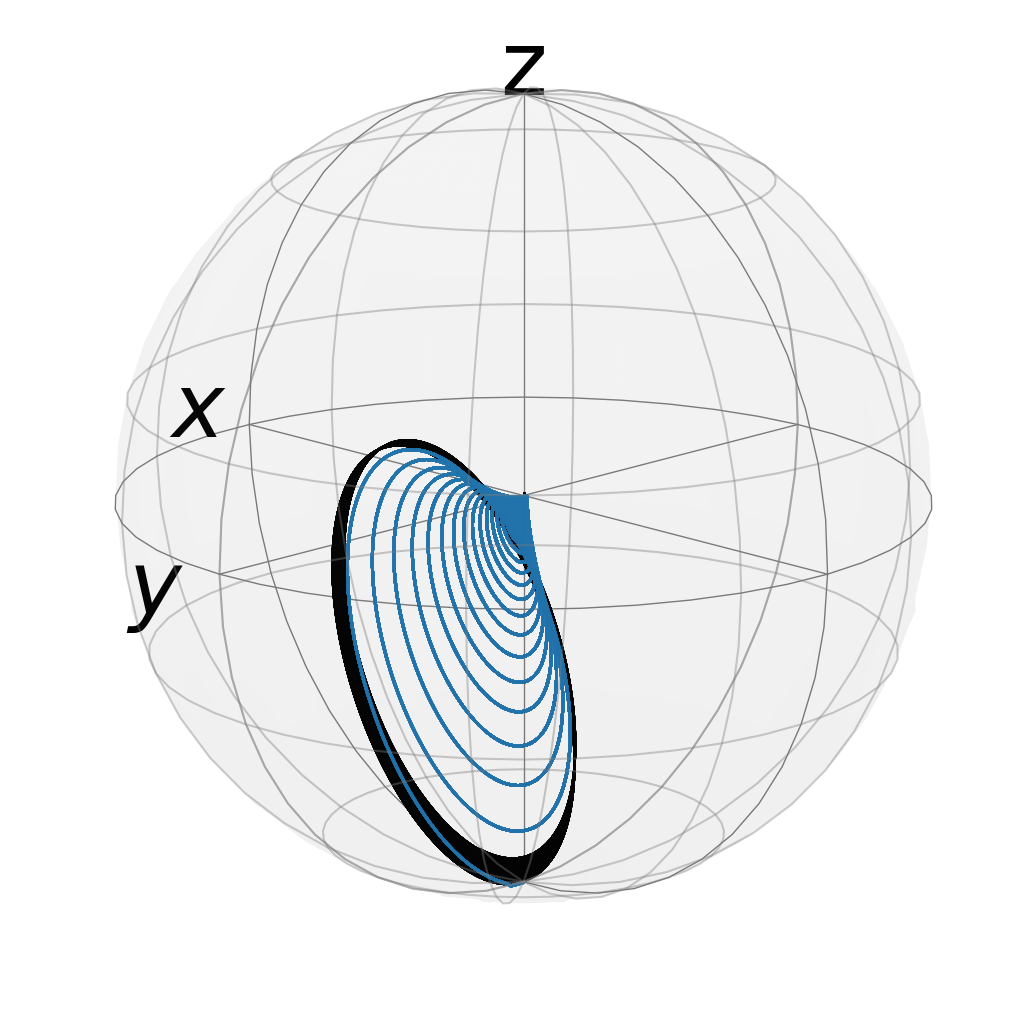}
\raisebox{3.8cm}{f)}\hspace*{-12pt}
\includegraphics[width=0.48\columnwidth]{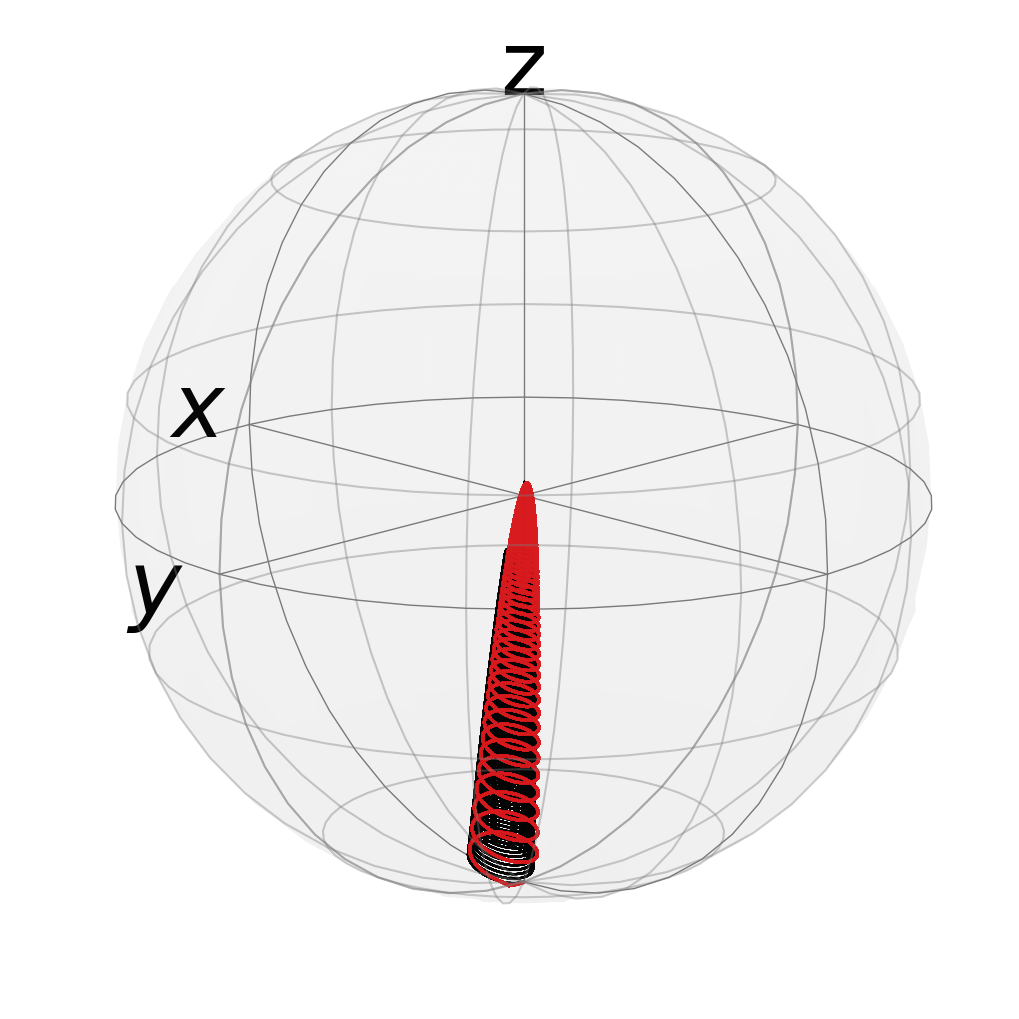}
\caption{Bloch sphere representation of the quantum dynamics, comparing the analytical (black) and  $L=16$ numerical (blue or red) results.
(a,b) referring to data in \Fig{fig:Nt-agree} with the same color, (c,d) to \Fig{fig:Nt-diff}(a), and (e,f) to  \Fig{fig:Nt-diff}(b). The figures were produced using QuTiP~\cite{johansson20121760}.
}
\label{fig:bloch-diff}
\end{figure}

\begin{figure}[t]
\includegraphics[width=\columnwidth]{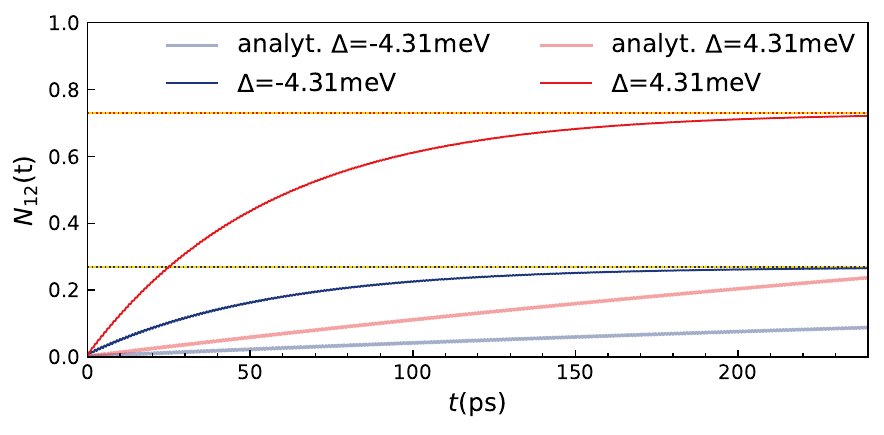}
\caption[]{
As in \Fig{fig:Nt-diff} with $g_F=262$~$\mu$eV but for a larger detuning $\Delta=\pm k_B T$ and $T=50$\,K. 
}
\label{fig:Nt-agree-2}
\end{figure}

This Appendix provides more examples of the dynamics of the reduced DM given by \Eq{P-full-before-LCE-zerotau}. Figure \ref{fig:bloch-diff} displays the dynamics of the results shown in \Fig{fig:Nt-agree} and \Fig{fig:Nt-diff} as trajectories on the Bloch sphere.

It also provides some more illustrations of the quantum dynamics for a larger detuning than used in the main text, as well as different QD sizes and deformation potentials.

Increasing the detuning to  $\pm\Delta= k_B T\approx 4.31$\,meV at $T=50$\,K for the system in \Fig{fig:Nt-agree}, the resulting dynamics is shown in \Fig{fig:Nt-agree-2}.
Clearly, at this large detuning (compared to the coupling), the component $D_-$ in the QD eigenstates is small, and accordingly small is the initial oscillation amplitude. The dynamics is dominated by a relaxation towards equilibrium, which is much faster for the exact result than for the analytical model. 
Both models give equilibrium values of 0.269 (0.731) for $\Delta =-k_B T$~$(k_B T)$ respectively.

\begin{figure}[t]	
\includegraphics[width=\columnwidth]{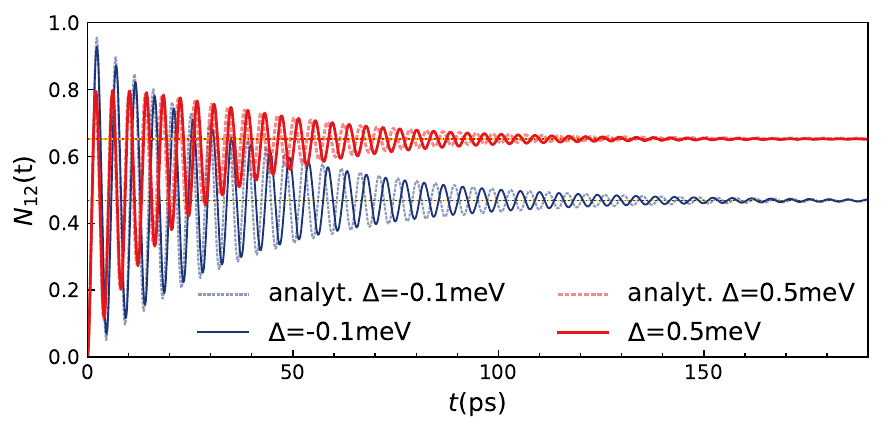}
\caption{As \Fig{fig:Nt-agree} but for $l=1.9$\,nm,  $D_c-D_v=-6.5$\,eV  $T=8$\,K,  $d=3.8$\,nm, and  $\Delta =-0.10$ and 0.50\,meV.
}
\label{fig:agree-better}
\end{figure}

\begin{figure}[t]	
\includegraphics[width=0.49\columnwidth]{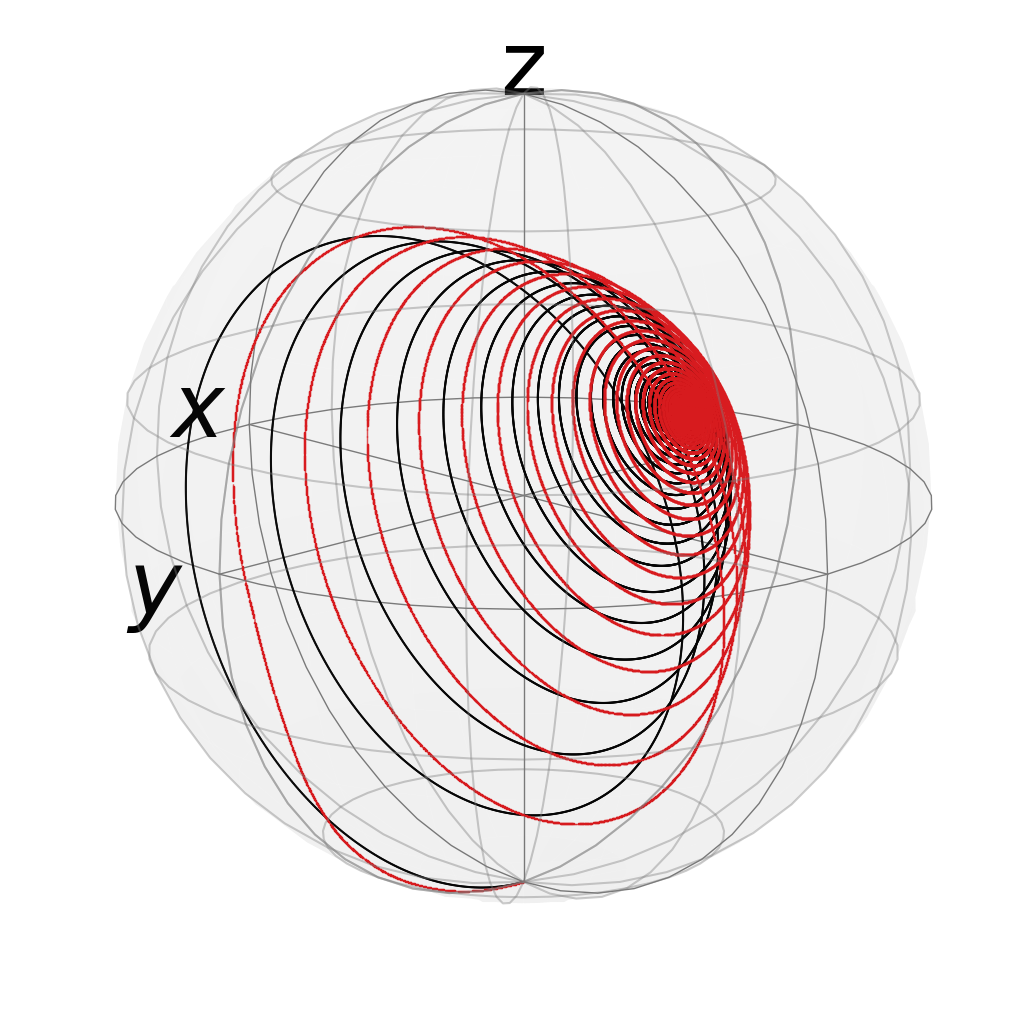}
\includegraphics[width=0.49\columnwidth]{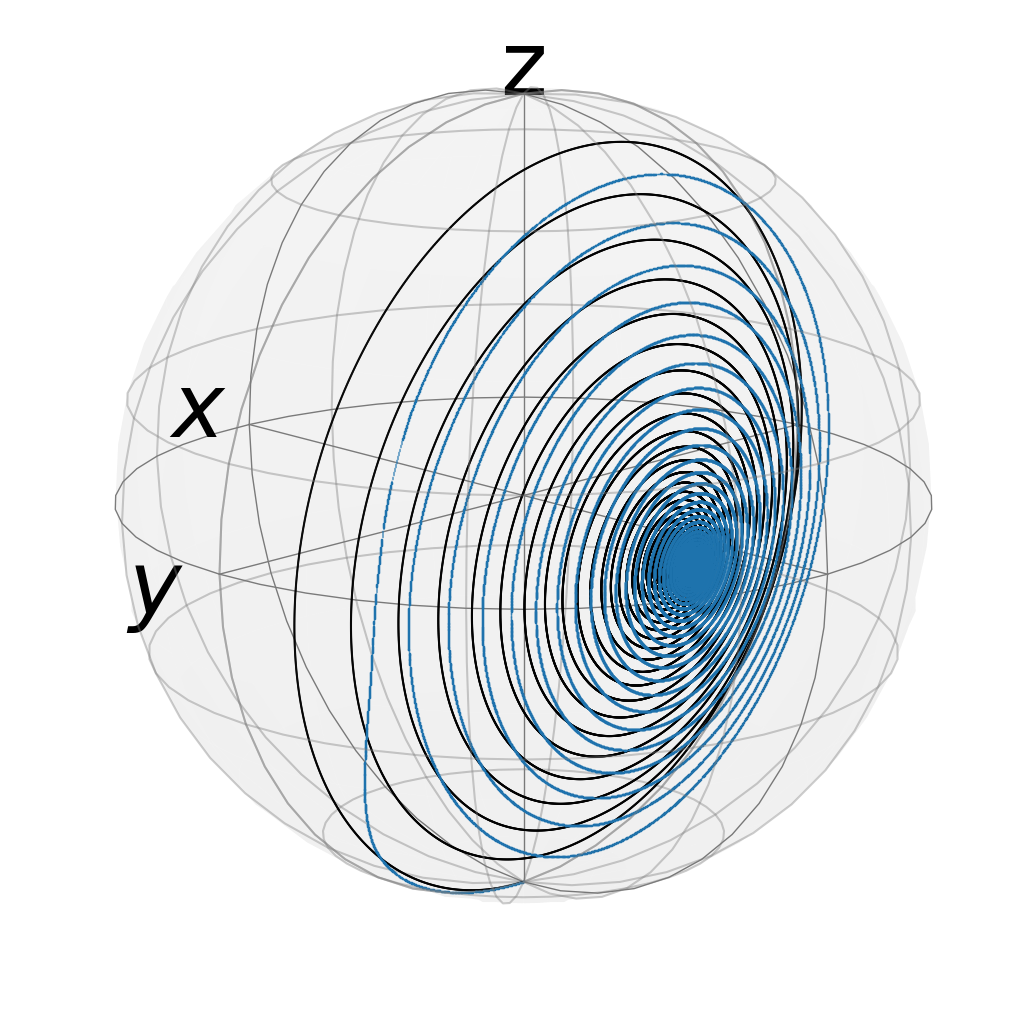}
\caption{Bloch sphere representation corresponding to results in \Fig{fig:agree-better}.}
\label{fig:agree-better-sph}
\end{figure}

To demonstrate a case with better agreement with the analytical model, we show in \Fig{fig:agree-better} the quantum dynamics for smaller QDs, using a smaller deformation potential difference $|D_c-D_v|$ and hence weaker coupling to phonons,  with its trajectory on the Bloch sphere given in \Fig{fig:agree-better-sph}. While the amplitude of the oscillations in the population dynamics and the long-time values are rather well reproduced by the analytical result (although as before the disagreement increases with decreasing magnitude of detuning), the oscillation periods do not agree.

\section{Fit results}
\label{app:fit}
Here we provide more details on the fit results reported in the main text. The constrained and unconstrained fits are compared. In addition to dephasing rates shown in \Figs{fig:timescales}{fig:Ttimescales} and discussed in the main text, the behavior of the remaining fit parameters with varying $d$ and $T$ is demonstrated.

\begin{figure}[t]	
\raisebox{4.9cm}{a)}\hspace*{-10pt}\includegraphics[width=\columnwidth]{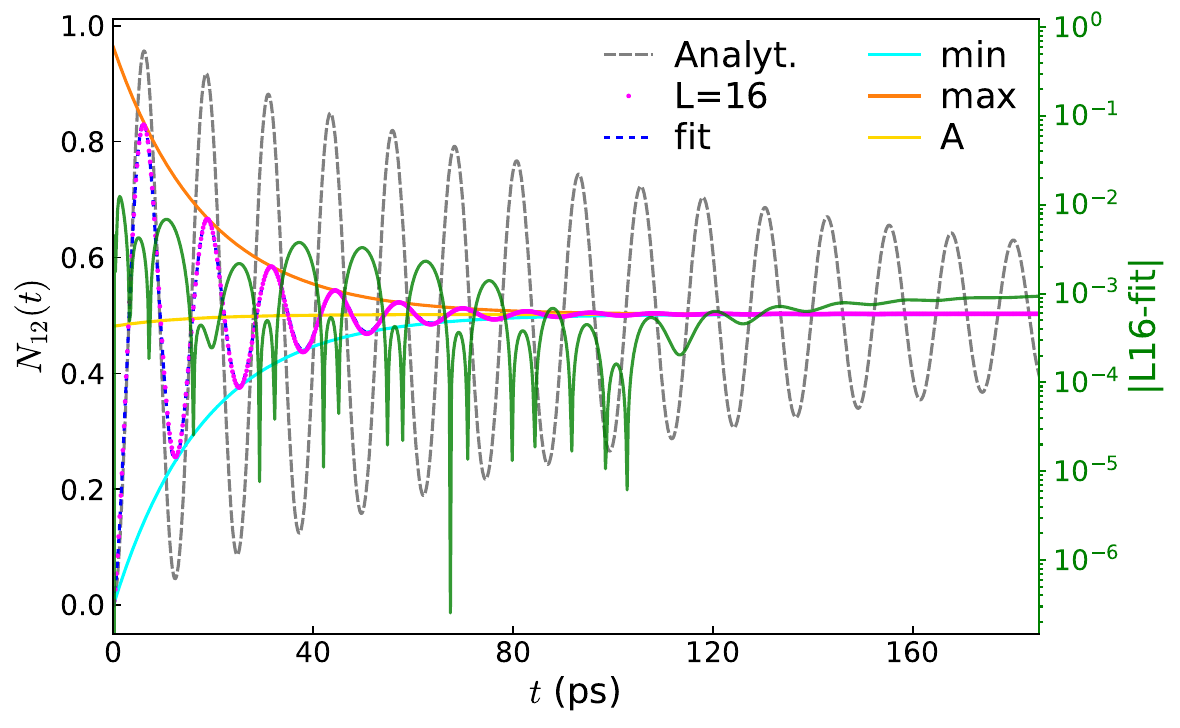}	
\raisebox{4.9cm}{b)}\hspace*{-10pt}\includegraphics[width=\columnwidth]{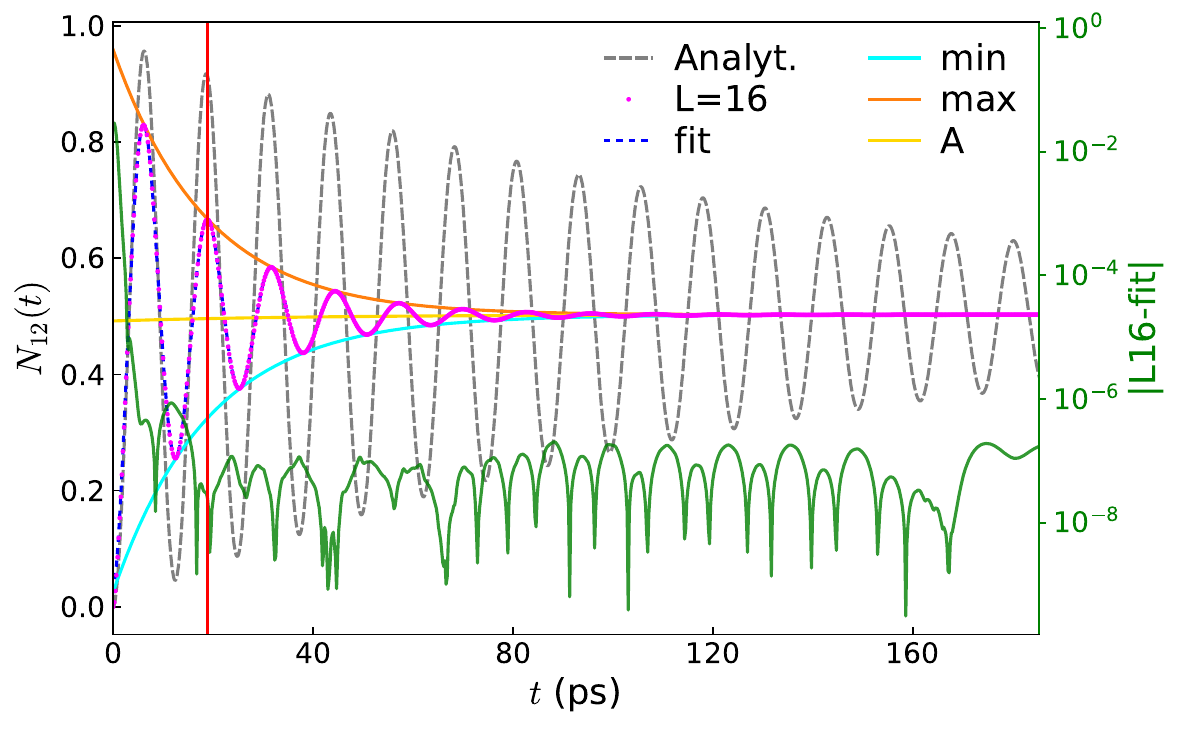}	
\caption{Fit results for $d=4.2$\,nm and other parameters as in \Fig{fig:timescales}, using a fit of the exact result, in (a) with $\mathtt{N}_{12}(0)=0$ constraint, in (b) without constraint. The green lines (right axes) show the absolute residual. The initial time range up to the vertical red line in (b) is not used for the unconstrained fit.}
\label{fig:tfit}
\end{figure}
Figure~\ref{fig:tfit} shows an example of both constrained and unconstrained fits applied to the same data. The absolute value of the residual is shown in green.
The unconstrained fit in \Fig{fig:tfit}(b) uses only the data for times larger than 19\,ps, and has a residual below $10^{-6}$ over the fitted time range, showing its excellent quality. This confirms that the analytical function represents the exact result well, giving confidence to the extracted parameters.
Notably, at early times, the fit error is more significant, increasing up to a few \%, reflecting the non-Markovian initial dynamics not captured in the analytical model.
The choice of the time range excluded from the fit is determined by the population decay characteristic timescale $T_1$, first extracted from the analytic model and then from the fitted numeric data as the fit is further refined.

The constrained fit, which uses the full time range starting from $t=0$, has instead a residual of the order of $10^{-3}$, showing that while it is trying to capture the non-Markovian initial dynamics, it sacrifices the accuracy of the long-term dynamics.
The observation of negative pure dephasing rates in \Fig{fig:timescales}(a) can be understood as a partial compensation in the fit of the initial non-Markovian dynamics at the cost of the subsequent Markovian dynamics.

\begin{figure}	
\raisebox{3.7cm}{a)}\hspace*{-8pt}\includegraphics[width=\columnwidth]{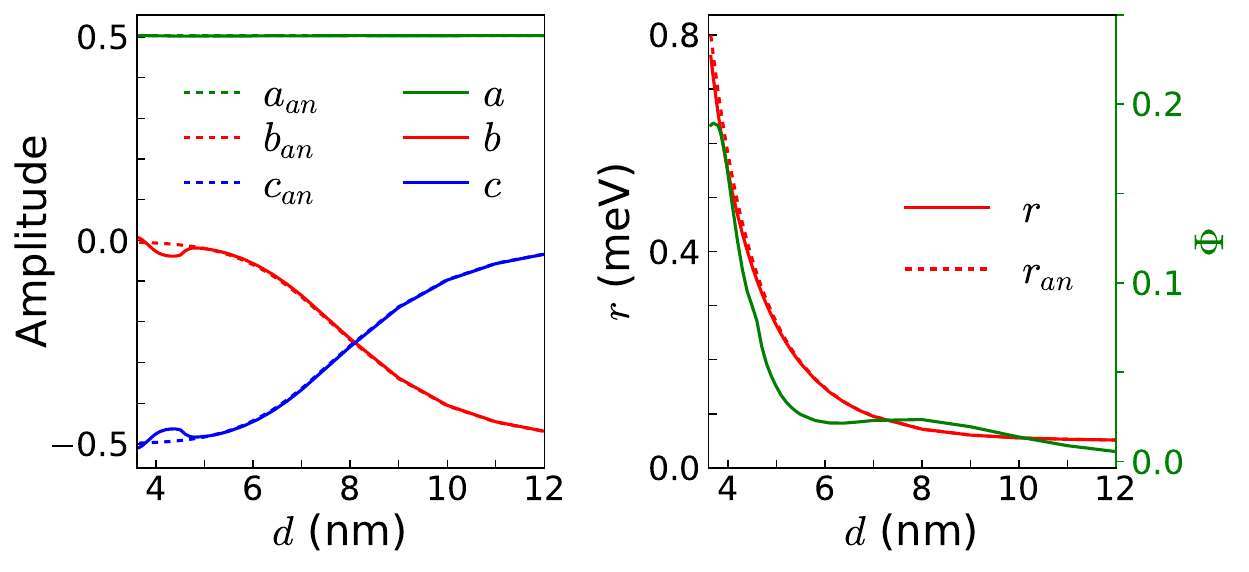}	
\raisebox{3.7cm}{b)}\hspace*{-8pt}\includegraphics[width=\columnwidth]{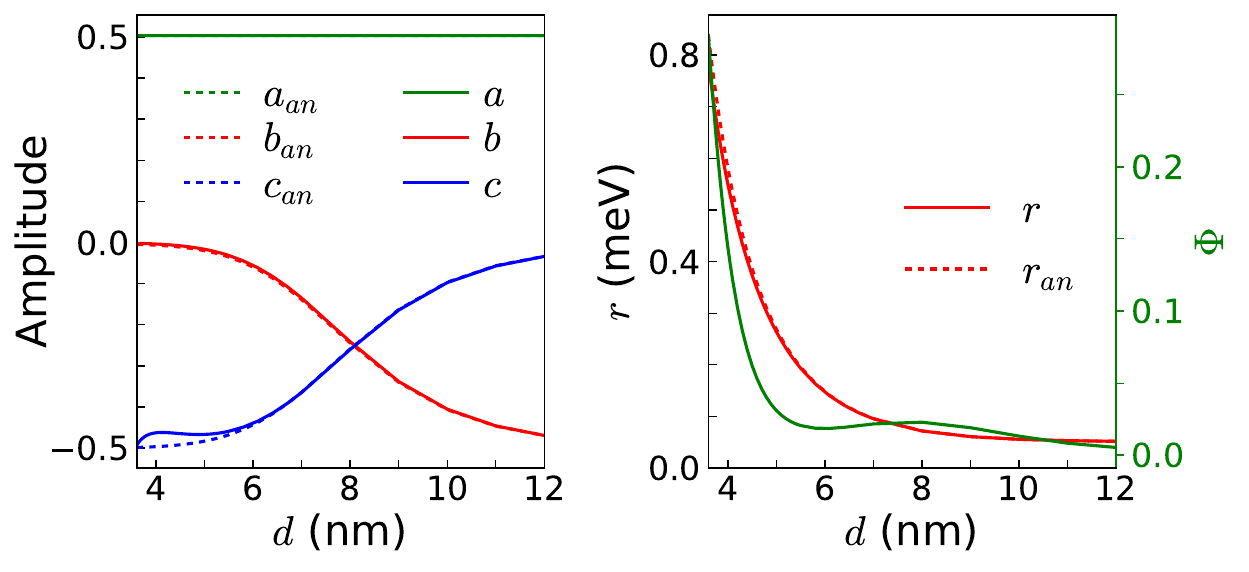}
\caption{As \Fig{fig:timescales} but showing the remaining fit parameters $a$, $b$, $c$, $r$ (solid lines), and $\Phi$ (right axis, green lines) in \Eq{Nfit} using a fit in (a) with $\mathtt{N}_{12}(0)=0$ constraint, and in (b) without constraint. For comparison the corresponding parameters $a_{\rm an}$, $b_{\rm an}$, $c_{\rm an}$, and $r_{\rm an}$ of the analytical result \Eq{N12analyt} are shown as dashed lines. 
}
\label{fig:fitphi}
\end{figure}

\begin{figure}
	\includegraphics[width=\columnwidth]{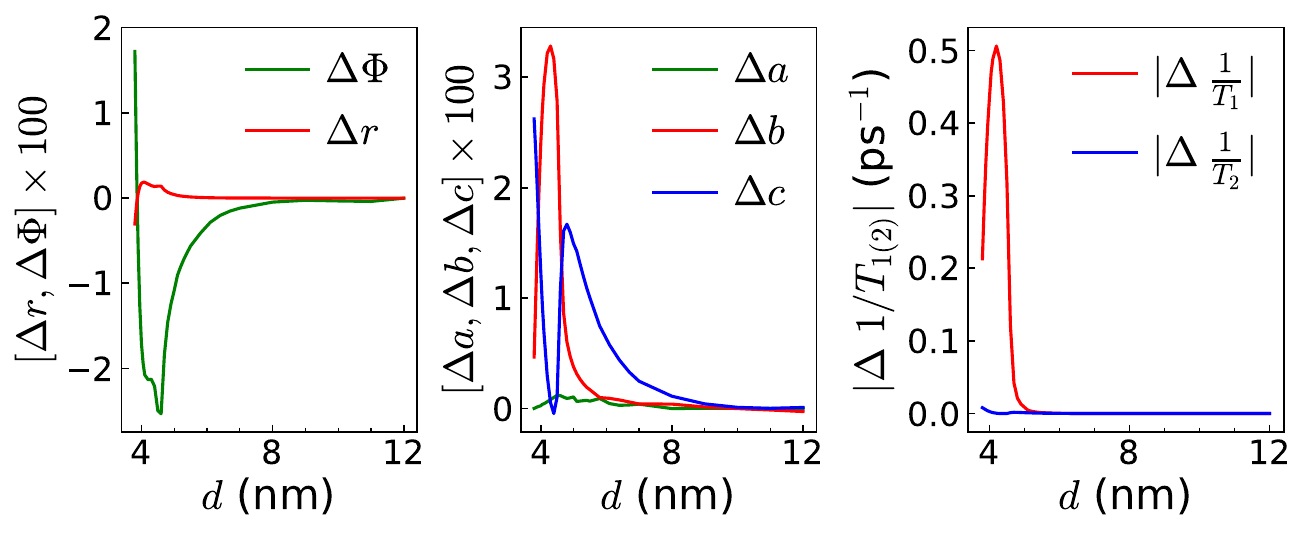}	
	\caption[]{Difference of the fit parameters between constrained and unconstrained fits versus dot separation $d$, referring to the same system as in \Fig{fig:fitphi}.
	}
	\label{fig:fitdiff}
\end{figure}

The additional parameters of the fit \Eq{Nfit} further to the decay rates shown in \Fig{fig:timescales} are given in \Fig{fig:fitphi}. The parameters $a$, $b$, $c$, and $r$, extracted from the exact solution closely follow the analytical result, but $b$ and $c$ deviate slightly at small $d$. The overall good agreement is surprising, given the significant deviation in the timescales $T_1$ and $T_2$ and the phase $\Phi$, which is absent in the analytical result. The equilibrium density parameter $a$ is essentially independent of distance in both constrained and unconstrained fits as expected.  The decay amplitude $b$ and the oscillation amplitude $c$ add up to an essentially constant value in the constraint fit shown in \Fig{fig:fitphi}(a) and both deviate for small distances from the analytical result, while without the constrain in \Fig{fig:fitphi}(b) only $c$ deviates significantly from the analytical result.

\begin{figure}
\includegraphics[width=\columnwidth]{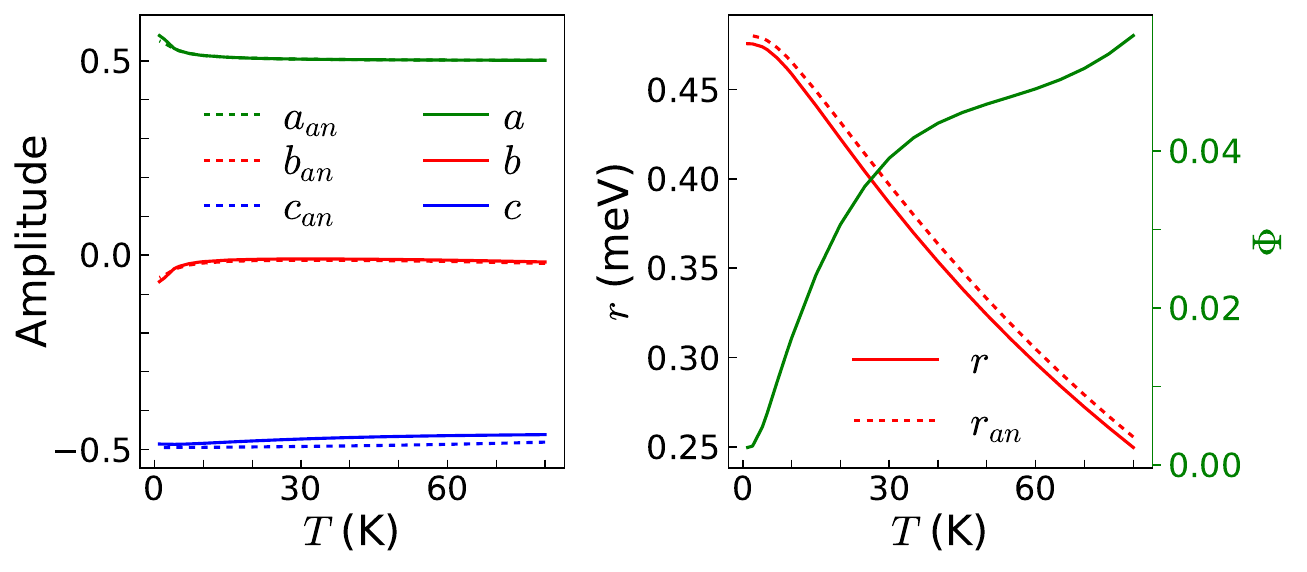}	
\caption{As \Fig{fig:fitphi}(b), but versus temperature for $d=4.7$\,nm.}
\label{fig:Tfitparams}
\end{figure}

Figure \ref{fig:Tfitparams} shows the additional parameters of the unconstrained fit versus temperature, further to the decay rates shown in \Fig{fig:Ttimescales}. The agreement of $a$ and $b$ with analytical result is excellent. For $c$ and $r$ it is again surprisingly good, given the disagreement in the dephasing rates.

\section{Thermal distribution and effective energy splitting}
\label{app:splitting}

Here we derive the thermal distributions \Eqs{N12inf}{N12inf0} and discuss
the differences between the long-time values of the population between the exact results and the analytical model. We also compare the effective splitting energies $R$ of the QD exciton hybrid states, with the thermal activation energy extracted from the long-time values of the population found analytically and numerically. 

Starting from the simpler case of uncoupled QDs, described by \Eq{N12inf0}, it is easy to see that, independent of the excitation, the final populations of QD1 and QD2 are proportional to $e^{-\beta\Omega_1}$ and $e^{-\beta\Omega_2}$, where the polaron shift can be also included by changing $\Omega_j\to \tilde{\Omega}_j $.  Then the normalized thermal distribution for QD1 is given by
\be
\mathtt{N}_{12}(\infty)= \frac{e^{-\beta\tilde{\Omega}_1}}{e^{-\beta\tilde{\Omega}_1}+e^{-\beta\tilde{\Omega}_2}}\,,
\ee
giving immediately \Eq{N12inf0}.

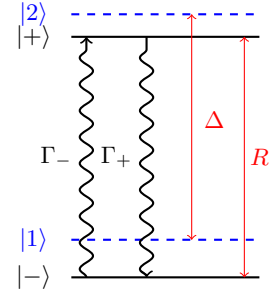
\begin{figure}[t]
\begin{tikzpicture}
\draw [thick] (-1,3.1) -- (1.5,3.1) node at (-1.5,3.1) {$|+\rangle$};
\draw [thick] (-1,-0.1) -- (1.5,-0.1) node at (-1.5,-0.1) {$|-\rangle$};
\draw [blue, thick, dashed] (-1,3.4) -- (1.5,3.4) node at (-1.5,3.4) {$|2\rangle$};
\draw [blue, thick, dashed] (-1,0.4) -- (1.5,0.4) node at (-1.5,0.4) {$|1\rangle$};
\draw[<->, red] (1.3,-0.1) -- (1.3,3.1) node at (1.5,1.5) {${R}$};
\tikzset{wiggle/.style={decorate, decoration=snake}}
\draw[<-, wiggle, thick] (0,-0.1) -- (0,3.1) node at (-0.4,1.5) {$\Gamma_+$};
\draw[->, wiggle, thick] (-0.8,-0.1) -- (-0.8,3.1) node at (-1.2,1.5) {$\Gamma_-$};
\draw[<->, red] (0.6,0.4) -- (0.6,3.4) node at (0.9,2) {$\Delta$};
\end{tikzpicture}
\caption{
Sketch of the uncoupled bare exciton states (dashed blue) and the energy levels (black) of the hybridized QD system, which are dressed, i.e. polaron shifted. There is a renormalized splitting $R$  between the hybridized states.  Upwards and downwards phonon-assisted transitions at rates $\Gamma_-$, $\Gamma_+$, respectively, are also shown.
}
\label{FGR-level}
\end{figure}

\begin{figure}[t]
\centering
\raisebox{3.8cm}{a)}\hspace*{-10pt}
\includegraphics[width=\columnwidth]{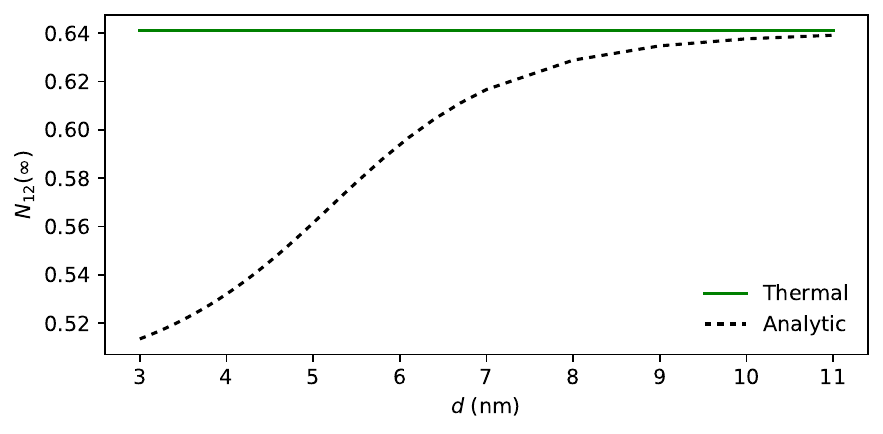}
\raisebox{3.8cm}{b)}\hspace*{-10pt}
\includegraphics[width=\columnwidth]{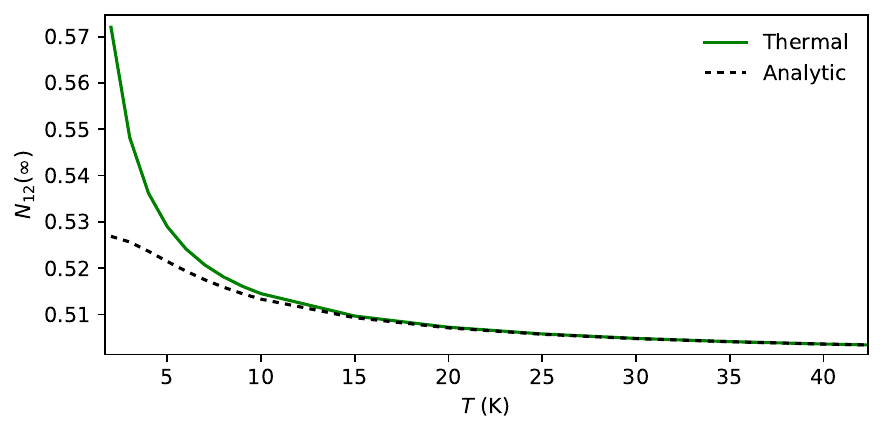}
\caption{
The long-time values predicted by the analytical result \Eq{N12inf} (black dashed lines) and the simple thermal distribution \Eq{N12inf0} (green lines) for (a) $\Delta=0.25$\,meV, $T=1$\,K; (b) $\Delta =0.25$\,meV, $d=3.8$\,nm and other parameters as in the main text.
}
\label{fig:long-time-val}
 \end{figure}

To derive \Eq{N12inf}, we use a similar logic but also take into account the hybridization of states \Eq{pmstates}, between which phonon-assisted transition, providing thermalization, occur, see \Fig{FGR-level} schematically illustrating these processes. Then the thermal population of states $|+\rangle$ and $|-\rangle$ is proportional to $e^{-\beta\Omega_+}$ and $e^{-\beta\Omega_-}$, respectively. The inverse transformation has the form
\be
|1\rangle= D_- |+\rangle + D_+ |-\rangle\,, \quad
|2\rangle= D_+ |+\rangle - D_- |-\rangle\,,
\label{states-inverse}
\ee
and therefore the thermal populations of states $|1\rangle$ and $|2\rangle$ are proportional to
$D_-^2e^{-\beta\Omega_+}+D_+^2e^{-\beta\Omega_-}$ and $D_+^2e^{-\beta\Omega_+}+D_-^2e^{-\beta\Omega_-}$, respectively. From this we obtain, by normalizing the distributions and using the identity $D_+^2+D_-^2=1$, the population of QD1:
\be
\mathtt{N}_{12}(\infty)= \frac{D_-^2e^{-\beta\Omega_+}+D_+^2e^{-\beta\Omega_-}}{e^{-\beta\Omega_+}+e^{-\beta\Omega_-}}
\,,
\ee
which is equivalent to \Eq{N12inf}.

Figure \ref{fig:long-time-val} shows a comparison of the long-time values given by the approximations \Eqs{N12inf}{N12inf0}. As already mentioned in \Sec{sec:analytics}, the two models agree well at high temperatures and large distances between the QDs, where the F\"orster coupling is small enough. In the opposite conditions, significant disagreements are seen.

The analytical approximation allows us to extract the relevant ``activation'' energy $r_{\rm ac}$ from the exact results, by using \Eq{N12inf} in the following way
\be
r_{\rm ac}=\frac{1}{\beta} \ln{\frac{1+\kappa}{1-\kappa}}\,,\quad
\kappa=\frac{r}{\Delta} (2 a-1) \,,
\ee
where long-time value $\mathtt{N}_{12}(\infty)$ is replaced with the fit parameter $a$, and $r$ is another fit parameter standing for the beat frequency in \Eq{Nfit}. In other words, we use the fit \Eq{Nfit} for determining the expansion coefficients $D_\pm$ contributing to \Eq{N12inf} in the form of the factor $\Delta/R$ and then resolve \Eq{N12inf} for $R$ standing only in the thermal factor.

\begin{figure}[t]
\centering
\raisebox{3.8cm}{a)}\hspace*{-10pt}
\includegraphics[width=\columnwidth]{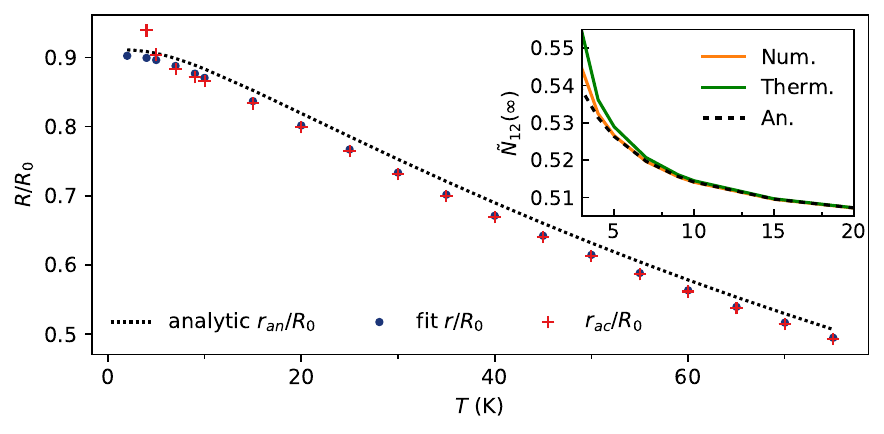}
\raisebox{3.8cm}{b)}\hspace*{-10pt}
\includegraphics[width=\columnwidth]{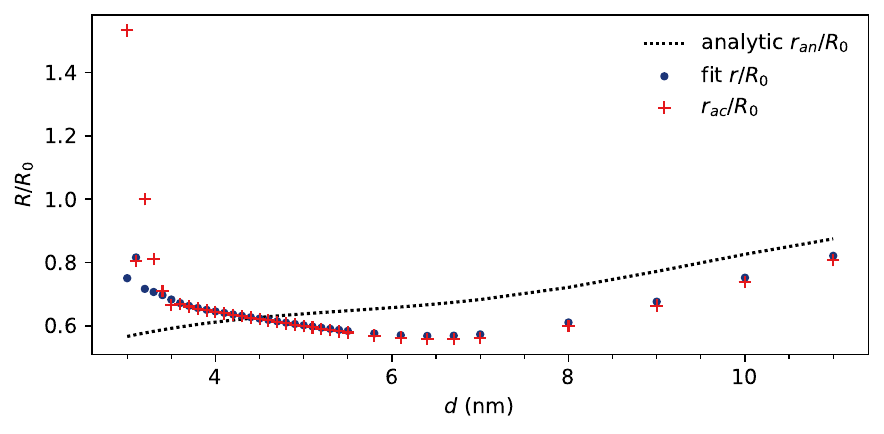}
\caption{
Effective splitting energy ratios, as predicted by analytical $R$ (black dashed), fitted numerical $r$ (blue dots) and the activation energy $r_{ac}$ (red crosses). The results are expressed as ratios with respect to the nominal Rabi splitting $R_0=\sqrt{\Delta^2+4g_F^2}$ without phonon interaction. The temperature dependence (upper panel) and the dependence on dot separation (lower panel) are shown.
The results are shown for (a) $d=4.7$\,nm; (b) $T=50$\,K; and other parameters as in the main text. 
}
\label{fig:energies-T-d}
 \end{figure}

We compare in \Fig{fig:energies-T-d}  four different values of the Rabi splitting: The analytical value $R$ defined in \Sec{sec:analytics}, the fitted value $r$, the activation energy $r_{\rm ac}$, and the nominal splitting $R_0=\sqrt{\Delta^2+4g_F^2}$. For a specific distance of $d=4.7$\,nm, the temperature dependence in \Fig{fig:energies-T-d}(a) shows a good agreement between the first three splittings and a significant offset from $R_0$ which increases with temperature. This is easy to understand in terms of the Huang-Rhys factor renormalization of the F\"orster coupling, which is explicit in the analytical model and strongly depends on temperature, see \Eq{HRg} within which $S$ itself increases with $T$. 

Notably, the analytic result is slightly above the other two results for the relative splitting in \Fig{fig:energies-T-d}(a).  The distance dependence in \Fig{fig:energies-T-d}(b) demonstrates a more significant disagreement between the analytical (black,dashed) and the other values, except a narrow region at around 4\,nm, although this is shown for a rather high temperature of 50\,K, where no agreement between analytical and numerical results is generally expected.

The observed excellent agreement between $r$ and $r_{\rm ac}$ supports the conclusion already made in \Sec{results:sep} that the generalization \Eq{Nfit} of the analytical model \Eq{N12analyt} is well suited to describe the population dynamics in the F\"orster transfer.

\section{Comparison with Ref.~\cite{Machnikowski2008quantum}}
\label{app:comparison}

Here we compare the exact solution with a high-quality approximation used in Ref.~\cite{Machnikowski2008quantum}.
While so far we have focused on spherical QDs, this Appendix also gives an example when the full calculation is applied to anisotropic QDs, as required for the comparison.

\begin{figure}[t]
\centering
\includegraphics[width=\columnwidth]{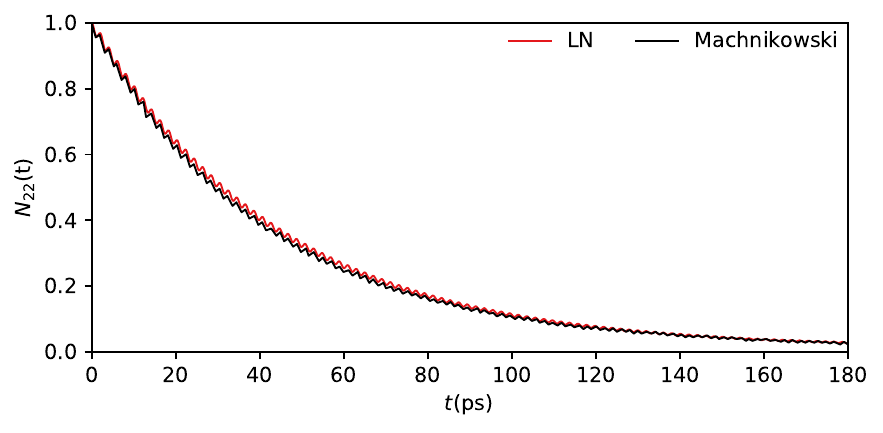}
\includegraphics[width=\columnwidth]{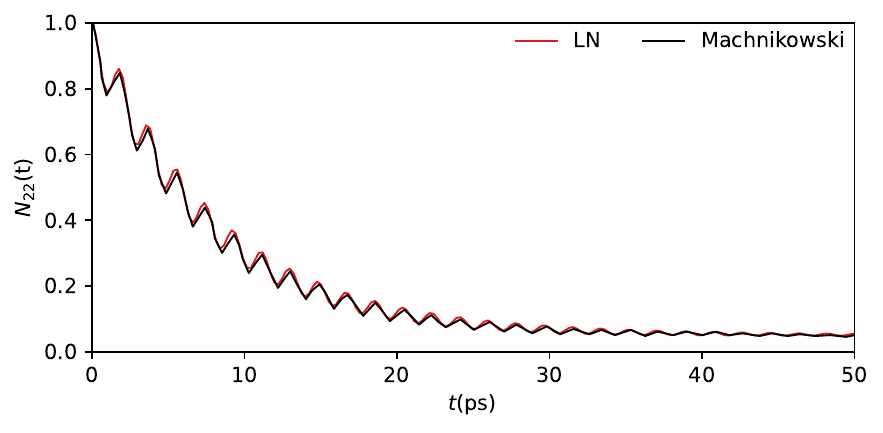}
\caption{
Comparison of the exact solution (red) to the correlation expansion (black) shown in Fig.~1~(c),(d) of Ref.~\cite{Machnikowski2008quantum}. The F\"orster coupling $g_F=0.2$\,meV (upper panel), $g_F=0.5$\,meV (lower panel). The other parameters are: $l_{\perp}=4.2$\,nm, $l_z=1.0$\,nm,  $D_c-D_v=-9.0$\,eV,  $T=4$\,K,    $d=6.0$\,nm,  $\Delta =$2\,meV.
}
\label{fig:machni}
 \end{figure}

Figure \ref{fig:machni} shows a comparison of the LN approach with a digitized data extracted from Fig.~1 of  Ref.~\cite{Machnikowski2008quantum}. The QDs are pancake shaped, with exciton confinement lengths $l_{\perp}=4.2$~nm  and $l_z=1.0$~nm, respectively, for in-plane and $z$-direction.

The overall agreement is very good, which is expected due to the high-quality calculation done in \cite{Machnikowski2008quantum}, which takes into account one- and two-phonon processes. The slight disagreement from our exact results may be due to the effect of the higher order processes neglected in Ref.~\cite{Machnikowski2008quantum} and also due to a little difference in the parameters used:  we have used the same in-plane confinement for the electron and hole $l_{\perp}=l_e=l_h=4.2$~nm, while Ref.~\cite{Machnikowski2008quantum} has taken them slightly different: $l_e=4.4$~nm and $l_h=4.0$~nm.

\bibliography{references}

\begin{thebibliography}{91}
\bibitem [{foo({\natexlab{a}})}]{footnote1}%
  \BibitemOpen
  \href@noop {} {} \ \bibinfo {note} {Strictly speaking, one
  should distinguish the detunings ${\Delta}={\Omega}_2-{\Omega}_1$ and
  $\tilde{\Delta}=\tilde{\Omega}_2-\tilde{\Omega}_1$, which become identical
  for the same-size quantum dots, due to equal polaron shifts.}\BibitemShut
  {Stop}%
\end{thebibliography}

\end{document}